\documentclass[]{jfm}
\usepackage{graphicx}
\usepackage[T1]{fontenc}
\usepackage{newtxtext}
\usepackage{newtxmath}
\usepackage{natbib}
\usepackage{array}
\usepackage{multirow}
\usepackage{multicol}
\usepackage{xcolor}
\usepackage{hyperref}
\hypersetup{
    colorlinks = true,
    urlcolor   = blue,
    citecolor  = black,
}

\newcommand{\RomanNumeralCaps}[1]
\linenumbers

\title{Stochastic Modeling and Upscaling of Hydrodynamic Transport in Geological Fractures}

\author{Alessandro Lenci\aff{1,2,3}
  \corresp{\email{alessandro.lenci@unibo.it}},
  Y. Méheust\aff{3,4},
   M. Dentz\aff{5}
 \and V. Di Federico\aff{1}}

\affiliation{\aff{1}Department of Civil, Chemical, Environmental, and Materials Engineering, Università di Bologna, Viale del Risorgimento 2, Bologna 40126, Italy
\aff{2}Department of Energy Science and Engineering, Stanford University, Stanford, CA 94305, USA
\aff{3}University of Rennes, CNRS, Géosciences Rennes -- UMR 6118, F-35042 Rennes, France
\aff{4}Institut Universitaire de France (IUF), France
\aff{5}Spanish National Research Council (IDAEA-CSIC), C. Jordi Girona 18-26, 08034 Barcelona, Spain}

\begin{document}
\newcommand{\vx}{{\mathbf x}}
\newcommand{\vu}{{\mathbf u}}
\maketitle

\begin{abstract}
Characterizing hydrodynamic transport in fractured rocks is essential for carbon storage and geothermal energy production. Multiscale heterogeneities lead to anomalous solute transport, featuring breakthrough curve (BTC) tailing and nonlinear growth of plume spatial moments. We focus on purely advective transport within synthetic geological fractures with prescribed relative closure $\sigma_a/\langle a \rangle$ and correlation length $L_\mathrm{c}$. We adopt a stochastic approach with multiple fracture realizations for each set of geometric parameters. Steady-state depth-averaged Stokes flow is solved under the lubrication approximation. Flow heterogeneity is organized over the correlation length $L_\mathrm{c}$. The ensemble-averaged velocity PDFs are insensitive to $L_\mathrm{c}$ but strongly influenced by $\sigma_a/\langle a \rangle$, particularly their low-velocity power-law scaling. A time-domain random walk (TDRW) simulation is used to compute plume spatial moments and outlet BTCs. The mean longitudinal plume position scales linearly with time at both early and late stages. The variance shows ballistic scaling at early times and a late-time behavior controlled by the low-velocity power law of the velocity PDF, with exponent $\alpha$ strongly influenced by $\sigma_a/\langle a \rangle$. The properties of the BTCs are also controlled by $\alpha$, including the broadening of the peak as $\sigma_a/\langle a \rangle$ increases, and the scaling of the power-law tails. Advective transport is also modeled using a one-dimensional continuous-time random walk (CTRW) that relies only on the velocity PDF, flow tortuosity, and flow correlation length. The CTRW reproduces the TDRW results and provides analytical predictions for the asymptotic transport scalings.
\end{abstract}


\section{Introduction}
\label{sec:intro}

Quantitative characterization of the subsurface flow and transport processes is fundamental to predict the hydrodynamic transport of solute or heat in numerous applications, such as underground carbon storage \citep{Hyman2019,Wang2023}, deep geothermal energy production \citep{Yoo2021}, underground nuclear waste disposal \citep{Poinssot2012,Hadgu2017,Tran2021}, and groundwater management and remediation \citep{Neuman2005}. 

Geological media, porous or fractured, are characterized by structural heterogeneity of hydraulic parameters at different scales. The hydraulic conductivity \( K \) of such formations can vary spatially abruptly and by several orders of magnitude: for clay or granite, $K$ is of the order of $10^{-12}~\textrm{m/s}$, and can reach $1~\textrm{m/s}$ for coarse sand or gravel \citep{Bear1972,SanchezVila2006}. In fractured formations, the genesis of fractures is attributable to tectonic activities, exhumation-induced temperature changes, or artificially induced reservoir stimulation \citep{Gale2014}. Fractures tend to develop into networks, whose permeability is orders of magnitude higher than that of the host rock matrix. Therefore, the flow occurs predominantly through such networks of interconnected fractures, and is governed by both the transmissivity of individual fractures and the network's connectivity and geometrical properties \citep{Long1985, bourWRR98,deDreuzyWRR2002,deDreuzyJGR2012,Viswanathan2022}. The former strongly depends on the pore-scale heterogeneity of the rough walls that constitute each fracture, while the latter determines network-scale flow patterns and the bulk volume of fluid conducted through pathways in the rock mass \citep{bourWRR98,Jing2007}. 

At the scale of a single fracture, flow takes place in the void space existing between the two rough walls. Since they were originally formed by fracturing of a rock mass, the fracture walls are self-affine \citep{Schmittbuhl1995a,Bouchaud1997}, that is, they exhibit peculiar scale invariance and spatial correlation properties. The two walls are geometrically matched, but only up to a characteristic correlation length, which implies that the aperture field is self-affine only below that scale \citep{Brown1995}. 
The aperture field is thus characterized by strong heterogeneity, with areas with wide apertures and low flow resistance contrasting with zones with low apertures, or contact zones that hinder the flow, associated with large viscous energy losses. 
 The flow through geological fractures has been studied for nearly forty years, using flow models either based on the depth-averaging over local apertures \citep{Brown1987,Meheust2001} or on full three-dimensional (Navier-)Stokes flow resolution \citep{brushWRR2003}. They have shown that aperture field heterogeneity results in flow channeling \citep{Brown1987,Meheust2001}, that is, the existence of preferential flow paths up to the scale of the correlation length \citep{Meheust2003}, which strongly impacts the fracture's permeability. This flow heterogeneity is all the larger as the relative fracture closure $\sigma_a / \langle a\rangle$, defined as the ratio of the aperture field's standard deviation to its mean value, is larger. Flow heterogeneity is even more pronounced for shear-thinning fluids, which exhibit stronger flow localization than Newtonian fluids \citep{Lenci2022b,Lenci2022a}. Besides, due to the intrinsically stochastic nature of fracture geometries, and the dominant impact on the flow of the largest Fourier modes of the aperture field \citep{Meheust2001}, a population of fractures with identical statistical geometrical parameters exhibits a very large dispersion over the hydraulic behavior, and this all the more as the correlation length is large \citep{Meheust2003}, so that any generic behavior can only be obtained through a stochastic approach, considering a large number of fracture realizations \citep{Lenci2022b,Lenci2024}. In turn, this stochastic flow heterogeneity strongly impacts solute transport through geological fractures. 

Starting from the pioneering works of \cite{deJosselin1958} and \cite{Saffman1959}, significant efforts have been dedicated to systematically quantifying solute or heat transport in heterogeneous porous and fractured media through stochastic modeling \citep{Dagan1989,Gelhar1993,Rubin2003,Yeh_Khaleel_Carroll_2015}. In particular, the onset of anomalous (i.e., non-Fickian) dispersion has been a subject of debate: laboratory and in-situ observations showed its dependence on a reference scale \citep{Gelhar1992}, indicating that it arises as a direct result of medium heterogeneity \citep{Warren1964,Pickens1981,Silliman1987}. A unified theory explaining scale effects in both flow and transport based on the underlying fractal structure of the hydraulic conductivity field of a porous medium was proposed by Di Federico and Neuman \citep{DiFederico1997,DiFederico1998a,DiFederico1998b}. In low-heterogeneity porous media characterized by a single and finite integral scale, the macro-dispersion coefficient evolves in time in the pre-asymptotic regime and tends to a constant asymptotic Fickian value proportional to the variance of the logarithm of hydraulic conductivity \citep{Dagan1984}. 
Recent research has established that anomalous transport phenomena in fractured media are deeply connected to structural heterogeneities such as mean aperture variability and correlation length, and that these can be effectively captured through stochastic frameworks such as Continuous Time Random Walk (CTRW) models. For instance, \citet{Hyman2019} and \citet{Hyman2021} investigated how flow channeling and network-scale structure in three-dimensional discrete fracture networks control Lagrangian velocity distributions and transport signatures, linking transport behaviors to the underlying network topology. \citet{Edery2016} emphasized the role of fracture aperture orientation and heterogeneity within the network in shaping breakthrough curve (BTC) tailing. \citet{Sund2021} and \citet{Elhanati2024} extended these analyses to karst and fractured aquifers, showing how large-scale structural features and temporal rainfall variability drive transport anomalies. 

For single fractures, early studies of solute transport have addressed the purely self-affine aperture field \citep{Ro-Pl-Hu,Pl-Hu-Ro-Ko,drazerPhysRevLett2004}, focusing on the link between the geometry of the solute front and that of the fracture. More recently, mostly numerical studies \citep{thompsonJGR1991, wang2014} have addressed solute transport in single fracture geometries. 
\cite{wang2014} successfully adopted a CTRW approach to reproduce non-Fickian transport behaviors from experiments by \cite{cardenas2007}, such as early solute breakthrough and heavy tailing in breakthrough curves. They found that predictions from the standard advection-dispersion equation became increasingly inaccurate when closing the fracture, while the CTRW model provided significantly better fits to the BTCs.

Stochastic transport in fractures has also been addressed in studies based on heterogeneous transmissivity fields, including those of \citet{Fiori2015} and \citet{Cvetkovic2014b}. In these works, transmissivity heterogeneity was modeled through multi-Gaussian random fields with prescribed variance and spatial correlation structure. These and other works on transport in highly
heterogeneous multi-Gaussian conductivity fields (most of them addressing porous media rather than fractures) \citep{Fiori2007,LBDC2008:2,Gotovac2009,Comolli2019} have identified
non-Fickian transport signatures characterized by early breakthrough and power-law BTC tailing. These transport signatures have also been upscaled using stochastic modeling approaches based on time-domain and continuous-time random walks \citep{delay2002,Berkowitz2006,Cvetkovic2014a,Noetinger2016}.

Given the experimentally documented self-affine nature of natural fracture walls \citep{Bouchaud1997,Schmittbuhl1995a}, an alternative and physically grounded description consists in deriving transmissivity variability directly from the fracture geometry. In this framework, velocity statistics emerge from aperture fluctuations through the Reynolds equation governing flow in rough fractures. The present study adopts this approach to 
investigate how the fracture's two main geometric parameters, the relative closure $\sigma_a/\langle a \rangle$ and the correlation length $L_\mathrm{c}$, control the velocity distributions and the associated purely-advective (i.e., infinite Péclet) transport behavior, under Stokes (i.e., creeping) flow conditions. We perform a stochastic analysis to account for the potential strong variability among the solute
transport behavior of fracture realizations with identical geometrical parameters. We investigate pre-asymptotic non-Fickian transport, and discuss the approach of the asymptotic Fickian behavior. We also investigate the impact of initial conditions (i.e., flux-weighted or uniform injection). To our knowledge, this is the first study that investigates the impact of the mean fracture aperture (or, equivalently, mean transmissivity) on advective transport considering a transmissivity field that relies  both on self-affinity of the aperture field below the correlation length and on a Gaussian probability density function of the local apertures. It is also the first study of transport in geological fractures that adopts a stochastic approach in which many fractures described by the same statistical geometrical parameters are considered, with results being interpreted in terms of both the mean behavior of the population and the fluctuations of the statistics around that mean. Because flow channeling in a rough fracture is sensitive to the particular realization of the aperture field, considering a single fracture does not, in general, provide 
a behavior that is representative of the mean behavior of the population. This is particularly true when the correlation length is not much smaller than the fracture size, so that the sampled aperture structures are not representative of the ensemble.  Here, we show how fluctuations around the ensemble average decrease when the correlation length becomes very small with respect to the fracture size; this is the limit for which ergodicity can be considered to hold for any given fracture of the population. Hence, we do not postulate ergodic representativeness a priori for advective transport, but show how it appears in the limit of very small correlation lengths.

To model transport numerically, we employ two complementary approaches: a time-domain random walk (TDRW) that directly tracks fluid particles, and an upscaled one-dimensional spatial CTRW framework in which the fluctuating Lagrangian velocity series is represented by an Ornstein–Uhlenbeck process \citep{Dentz2016,Kang2017,Morales2017}. Such random walk approaches, which directly model particle motion in heterogeneous media, have been proposed to investigate non-Fickian behavior while avoiding numerical diffusion. Among RWs, TDRWs have been adopted to model particle motion, relying on evenly displacing particles along streamlines with variable residence times \citep{Noetinger2016}, as the flow field is organized on fixed length scales \citep{Berkowitz1997,Benke2003,Kang2011}. TDRWs are computationally efficient as they avoid unnecessary iterations in low-velocity zones. Continuous-time random walk (CTRW) upscaling formulations, on the other hand, allow linking transport scaling to the low-velocity structure of the flow field \citep{Painter2002,Comolli2019,Puyguiraud2019b,Hyman2021,Dentz2023}. The simulations show that, in single synthetic fractures with realistic wall roughness, the interplay of relative closure and correlation length gives rise to a wide range of transport regimes, even in the high-Péclet limit. 
Moreover, these behaviors can be efficiently captured using the upscaled CTRW-based Ornstein–Uhlenbeck model, validating its application at the fracture scale and confirming its robustness to link fracture geometry to macroscopic transport scaling, under moderate ergodicity conditions, from the sole knowledge of the velocity statistics, flow tortuosity, and a Lagrangian correlation length. 
In addition, the CTRW model allows us to  theoretically predict the exponents of long time power law exponents in the breakthrough curves from the low velocity asymptotic behavior of the Eulerian velocities' probability density function.

The organization of the article is as follows: Section \ref{sec:flow} describes rough fracture geometries and how synthetic fractures with such geometries can be generated numerically, the derivation of the Reynolds equation describing depth-averaged Stokes flow, and the finite-volume Stokes-flow solver. Section \ref{sec:transport} presents the models of hydrodynamic transport in heterogeneous media, which are a time domain random walk particle tracking scheme and a one-dimensional upscaled model based on a CTRW. Section \ref{sec:results} reports on the results of the stochastic analysis for the velocity field and advective transport. Section~\ref{sec:conclusions} summarizes the study and provides potential further prospects for future studies.

\section{Methods}
We investigate advective transport through synthetic rough-walled geological fractures. The modeling framework presented here relies on the following assumptions. The flow is assumed to be steady, incompressible, and isothermal, and is governed by the Stokes equations. Because of the small fracture aperture and the gradual spatial variation of the walls, we adopt the lubrication approximation, yielding the two-dimensional Reynolds equation. Fracture walls are modeled as self-affine and statistically isotropic, with their mutual matching controlled by a finite correlation length. Ideal plastic closure is used to model contact zones, i.e. negative aperture values are set to zero. Particle transport is modeled in the purely advective limit (i.e., at infinite Péclet number). The transported entities are fluid particles, which have no mass, follow streamlines, and interact neither with the solid boundaries nor with each other.

\subsection{Flow in Geological Fractures}
\label{sec:flow}
\subsubsection{Heterogeneous Aperture Fields}
\label{sec:aperture}
The aperture field $a$ of a geological fracture is defined as the distance between the two wall topographies along the direction perpendicular to the fracture’s mean plane. For a fracture whose mean plane is horizontal, if the top and bottom wall topographies (each with zero mean) are denoted respectively by $h_\textrm{u}$ and $h_\textrm{b}$, then the aperture field is
\begin{equation}
\label{eq:def_a}
    a(\vx) \;=\; h_\textrm{u}(\vx)\;-\;h_\textrm{b}(\vx)\;+\;a_\textrm{m},
\end{equation}
where the mechanical aperture $a_\text{m}$ is the distance between the mean planes of the wall topographies, and $\vx = (x_1,x_2)^\top$ is the position vector in the $(x_1,x_2)$ plane (with $\top$ denoting the transpose).

The fracturing process that forms the fracture causes the fracture walls’ topography to exhibit self-affine scale invariance across all scales \citep{Bouchaud1990,Schmittbuhl1995a,Candela2009}. One consequence of this self-affinity is that the two-dimensional (2D) power spectral density $\mathcal{G}$ of the wall topography scales as a power law of the wavenumber modulus $k$, namely:
\begin{equation}\label{eq:Fourier}
    \mathcal{G}\;\propto\; k^{-2(H + 1)},
\end{equation}
where $H$ is the so-called Hurst exponent \citep{Schmittbuhl1995a}. The value of $H$, governed by the fracturing process, typically lies in the range $[0.5, 0.9]$ and is nearly universal (close to 0.8) for most brittle materials, including igneous rocks such as granite and basalt \citep{Bouchaud1990,Meheust2000}. A value of 0.5 is usually measured in sandstone, due to its intergranular fracturing \citep{Boffa1999}. In any case, the Hurst exponent can be considered a petrophysical parameter of the geological formation.

The walls of a fresh brittle fracture are identical at all scales. In contrast, in geological fractures, due to the combined action (over geological times) of chemical weathering and interaction between the fracture walls, such as tectonics-induced relative movement and mechanical grinding, their topographies are only matched at scales larger than a characteristic matching scale that we denote the correlation length $L_\mathrm{c}$ \citep{Brown1995}. Consequently, the aperture field, being related to the two topographies through Eq.~\eqref{eq:def_a}, retains self-affine scaling only at scales smaller than $L_\mathrm{c}$. In particular, the power spectral density of the aperture field only exhibits a power law of the type presented in Eq.~\eqref{eq:Fourier} at wave numbers larger than $k_\mathrm{c} = 2\pi / L_\mathrm{c}$.

In this work, fracture aperture fields are thus generated according to the algorithm proposed by \cite{Meheust2001} and \cite{Lenci2022a}. Specifically, the algorithm enforces Eq.~\eqref{eq:Fourier} for wavenumbers $k > k_\mathrm{c}$, corresponding to scales smaller than the correlation length, and assumes no dependence on $k$ for $k \le k_\mathrm{c}$. 
This ensures that the aperture field is self-affine up to the correlation length $L_\mathrm{c}$ and uncorrelated beyond it. This is achieved by multiplying the 2D white-noise Fourier transform by
\begin{equation}
    \biggl[\max\!\Bigl(\sqrt{k_{x_1}^2 + k_{x_2}^2},\,k_\mathrm{c}\Bigr)\biggr]^{-2(H+1)},
\end{equation}
and then taking the inverse Fourier transform of the product to generate a 2D isotropic aperture field.  
Finally, the aperture field is rescaled and translated to match the desired mean, $\langle a\rangle$, and standard deviation, $\sigma_a$. If the relative fracture closure, defined as $\sigma_a / \langle a\rangle$, is sufficiently large, some regions of the generated aperture field may become negative; these are set to zero (ideal plastic closure). In such cases, the mean aperture $\langle a\rangle$ no longer coincides with the mechanical aperture $a_\textrm{m}$. Throughout the Results section, we therefore characterize fracture closure using $\sigma_a / \langle a \rangle$. This is the notation used in the Results section.
\begin{figure}
\centerline{\includegraphics[width=0.9\textwidth]{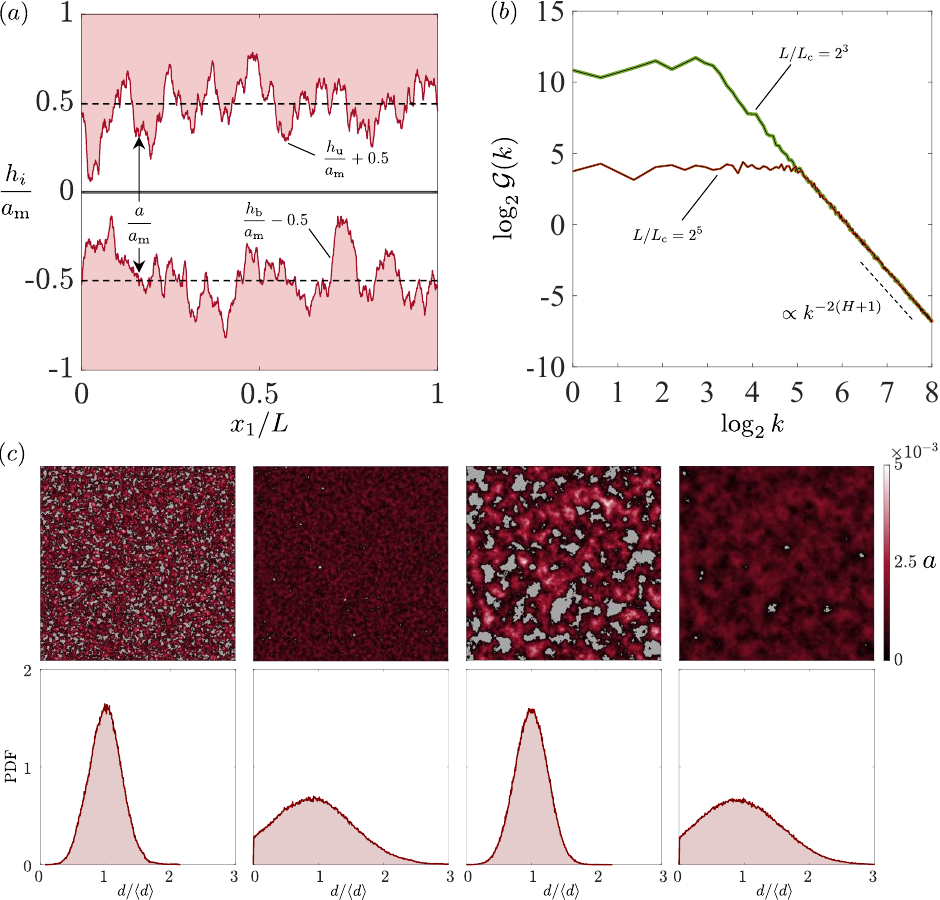}}
\caption{Overview of the geometries and associated spectra from synthetic geological fractures. Figure~\ref{Fig1}(a) shows one longitudinal profile of the fracture, with the representation of the walls and the definition of the geometrical fields that define the fracture geometry. Figure~\ref{Fig1}(b) presents the power spectrum for the cases $L/L_\mathrm{c} = 2^5$ in red and $L/L_\mathrm{c} = 2^3$ in green. Figure~\ref{Fig1}(c) shows four aperture field realizations arranged from left to right, each with its corresponding local aperture probability density function (PDF) displayed below. The first two realizations correspond to $L/L_\mathrm{c} = 2^3$ with closure values of 0.25 and 0.75, respectively, while the last two correspond to $L/L_\mathrm{c} = 2^5$, also with closures of 0.25 and 0.75. In all cases, the local aperture PDFs are approximately Gaussian, with a cutoff at zero when contact regions are present in the fracture plane.}
\label{Fig1}
\end{figure}
Figure~\ref{Fig1} provides an illustration of fracture geometries and their corresponding spectral characteristics. Figure~\ref{Fig1}(a) presents one longitudinal profile of a synthetic rough fracture, where the fracture walls are depicted along with the key geometric fields that define the fracture geometry. In Figure~\ref{Fig1}(b), the power spectrum is shown for two different values of $L/L_\mathrm{c}$, showing the cutoff of the power law behavior at $k_\mathrm{c}$. Figure~\ref{Fig1}(c) shows aperture fields and corresponding probability density functions for closures of 0.25 and 0.75, each at two system sizes: $L/L_\mathrm{c} = 2^3$ and $L/L_\mathrm{c} = 2^5$. The generated aperture fields are self-affine up to $L_\mathrm{c}$, i.e., long range correlations up to that scale, which explains why the size of the largest areas of uniform colors is $L/L_\mathrm{c}$; the size of the closed regions also reflects the value of $L_\mathrm{c}$, for the same reason. Note that in the limit $L_\mathrm{c} \to 1$, the aperture field would approach uniformity. Additionally, the numerical discretization introduces an implicit small-scale cutoff at the grid size. 
The unresolved fine-scale roughness is, in reality, a continuation of the self-affine behavior down to micrometric length scales. Since small-scale spectral modes have diminishing influence on the flow organization, resolving the geometry more would not result in a significantly different velocity field. The adopted spatial resolution ($2^{10}$ control volumes per side) ensures that geometric features are accurately represented down to the grid scale, which is much smaller than the correlation length in all cases. Finally, the probability density function of the local apertures is approximately Gaussian in all cases, as shown in Figure~\ref{Fig1}(c), with a cutoff at zero when contact regions are present in the fracture plane.

Geological fractures typically have mean apertures ranging from tens of microns to several millimeters. In this work, we set $\langle a\rangle = 1$ mm, representative of typical fracture apertures in crystalline and sedimentary rocks, as indicated by various laboratory measurements \citep{Watanabe2008}. 
Furthermore, we consider two different values of the fracture relative closure $\sigma_a/\langle a\rangle$ to represent distinct degrees of aperture variability. We also fix the correlation length to $L_\mathrm{c} = 0.1\,\mathrm{m}$, based on characteristic lengths observed in fractured rock outcrops and laboratory studies \citep{Brown1995,Meheust2000}. We then vary the fracture length $L$, so as to obtain different values of the ratio $L/L_\mathrm{c}$. This means that we consider the correlation length to be a property of the fracturing process, which is uniform over a given fractured medium; fractures of different lengths within that medium still exhibit the same correlation length $L_\mathrm{c}$.

\subsubsection{Flow Equation}
\label{sec:Reynolds}
Under steady and isothermal conditions, the viscous flow of an incompressible fluid through a fracture of variable aperture $a=a(x_1,x_2)$ is governed by the Stokes equations:
\begin{align}
\label{eq:conservation}
    \boldsymbol \nabla' P  &= \mu \boldsymbol \nabla'^2\mathbf{u}', && \boldsymbol \nabla'\cdot\mathbf{u}' = 0,
\end{align}
where $\mathbf{x}=(x_1,x_2,x_3)^\top$ is the position vector, $\mu$ is the fluid's dynamic viscosity, $\mathbf{u}'=(u_1,u_2,u_3)^{\top}$ is the velocity field, $P$ is the pressure field (including gravity effects), and $\boldsymbol \nabla'$ is the gradient operator. If the aperture field $a$ is sufficiently smooth, we apply the lubrication approximation, according to which the in-plane component of the velocity field is much larger than its out-of-plane component. It follows that the gap-averaged velocity field, $\mathbf u= (u_1,u_2)^{\top}$, satisfies the Darcy-type equation \citep{Zimmerman1996}
\begin{align}
\label{eq:darcy}
\vu = - \frac{a^2}{12 \mu} \boldsymbol \nabla P, 
\end{align}
at any position within the mean fracture plane and with $\boldsymbol\nabla$ denoting the gradient operator in the $(x_1,x_2)$ plane. Since the three-dimensional velocity field $\mathbf{u}^\prime$ is divergence-free, 
the depth-integrated velocity $\mathbf{q} = a\,\mathbf{u}$, previously referred to as the local flux~\citep{Meheust2001}, 
also satisfies the continuity condition $\boldsymbol{\nabla}\!\cdot\!\mathbf{q} = 0$. By combining this condition and Equation~(\ref{eq:darcy}), 
one then obtains the Reynolds equation governing the pressure distribution within the fracture:
\begin{equation}\label{eq:Reynolds}
  \boldsymbol \nabla \cdot (a^3 \boldsymbol \nabla  P) =0. 
\end{equation}
We determine the probability density function (PDF) of Eulerian flow velocities $u_e(\vx) = \|\vu(\vx)\|$ by spatial sampling in the flow domain $\Omega$ of area $|\Omega|$:
\begin{align}
\label{eq:fe}
f_e(u) = \frac{1}{|\Omega|} \int_{\Omega} d\mathbf{x}\; \delta[u - u_e(\mathbf{x})].
\end{align}
Furthermore, we calculate the advective tortuosity, which is given by 
\begin{align}
\label{eq:tortuosity}
\chi = \frac{\langle u_e \rangle}{\langle u_1\rangle},
\end{align}
where the angular brackets denote spatial averaging. The advective tortuosity $\chi$ quantifies how much the average particle trajectories deviate from the mean flow direction due to flow channeling. It measures the geometric elongation of transport paths and is critical for relating streamwise distances to longitudinal displacements in upscaled models.

\subsubsection{Numerical Implementation}
\label{sec:Numerical}
\begin{figure}
\centerline{\includegraphics[width=0.9\textwidth]{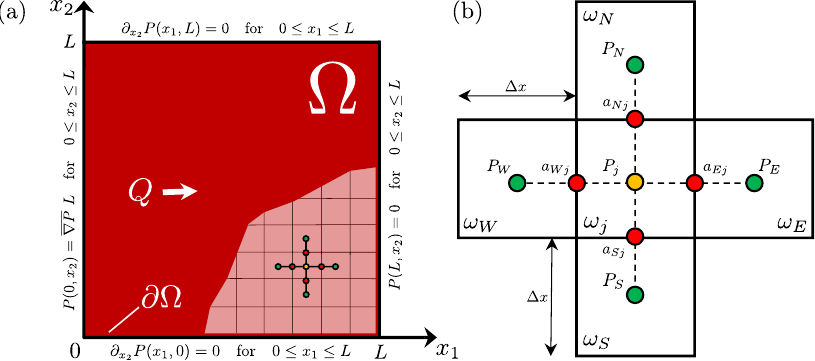}}
\caption{(a) Representation of the domain partitioning with boundary conditions. (b) Finite volume scheme 5-point stencil: pressure is defined at the centre of each control volume, while the local aperture is estimated along the edge of the cells by arithmetic averaging.}
\label{Fig2}
\end{figure}

The fracture is modeled as a two-dimensional square domain ($\Omega$) of side length $L$ and boundary $\partial\Omega=\partial\Omega_\textrm{D}\cup\partial\Omega_\textrm{N}$, as shown in Figure~\ref{Fig2}a). This domain corresponds to the projection of the fracture volume onto its mean plane, positioned at an equal distance between the mean planes of the two walls (which are parallel to each other). The flow equation~\eqref{eq:Reynolds} is solved numerically under the action of an externally imposed macroscopic pressure gradient $\overline{\nabla P}$. Along the left- and right-hand side of the fracture ($\partial \Omega_\textrm{D}$), Dirichlet boundary conditions are considered: $P(0,x_2)=\overline{\nabla P}\, L$ and $P(L,x_2)=0$, respectively. Neumann boundary conditions are imposed on the remaining portion of the boundary ($\partial \Omega_\textrm{N}$). The aperture field $a(\vx)$ is defined over a discrete regular partition of $\Omega$ in $n^2$ non-overlapping control volumes $\omega$ of linear size $\Delta x=L/n$, with $n=2^{10}$. This ensures that both the geometric heterogeneity and the resulting velocity gradients are well resolved; for instance, when $L_\mathrm{c} = 0.1$ m, and $L/L_\mathrm{c}=32$, the domain size $L$ is 3.2 m and that of each control volume is $\Delta x = 3.1$ mm, which is comparable to the mean aperture. Consequently, the convergence of the pressure and velocity fields is ensured, in good consistency with prior studies \citep{Meheust2001,Lenci2022a}. Figure~\ref{Fig2}(a) shows a schematic of the computational domain, the partitioning into finite volumes, and the boundary conditions applied. 
A finite-volume discretization of the Reynolds equation is formulated to solve the equation (as shown in Figure~\ref{Fig2}(b)). For each control volume $\omega_j$, whose set of neighboring control volumes is indexed as $\sigma(j)=\{N, S, E, W\}$, the following linear equation holds:
\begin{equation} \label{eq21}
\sum\limits_{k\in\sigma(j)} c_{k}^{(j)}(p_{k}-p_{j})=0
\end{equation}
where $p_i$ denotes the pressure in the centre of the $i$-th cell, and the coefficient $c_k^{(j)}=(a_k+a_j)^3/(8 \Delta x^2)$ is obtained as the arithmetic average of the local apertures of cells $k$ and $j$. It is the same in the $1$- and $2$- directions. Note that the harmonic mean would be energetically consistent and thus preferable in principle, but we chose to employ the arithmetic mean to mitigate numerical instability issues. Specifically, the arithmetic average is less sensitive to the ill-conditioning caused by the strong aperture variability, which can span several orders of magnitude \citep{mazzia2011bad}. The coordinate of the centre point of the pixel $i$ is denoted by $\vx^{(i)}$. 
Eq. \eqref{eq21} is equivalent to the linear system $\mathbf A \mathbf{p}=\mathbf{f}$, whose coefficient matrix $\mathbf A$ is: 
\begin{equation}\label{eq23}
A_{ij}=
\begin{cases}
-\sum\limits_{\substack{k\in\sigma(j)}}c_{k}^{(j)} \quad &\textrm{if} \quad i=j;\\
 c_{i}^{(j)}\quad &\textrm{if} \quad i\in\sigma(j);\\
0 \quad &\textrm{otherwise~,}
\end{cases}\, .
\end{equation}
while the vector $\mathbf f$ is
\begin{equation}
f_{j}= -c_{W}^{(j)}\overline{\nabla P}\, L, 
\end{equation}
at the nodes in the left boundary and zero otherwise. The system is well-posed provided that the coefficient matrix $\mathbf{A}$ is an M-matrix, which is ensured in this case by strictly positive diagonal entries $A_{ii}$. This condition requires the enforcement of a non-null aperture $a_0$ at the contact zones: the approximation does not affect the solution as long as $a_0$ is sufficiently small. 
To estimate velocities at the centre of each control volume, we use an arithmetic average of the edge velocities. While methods such as that of \citet{Pollock1988} are designed for accurate streamline tracing in heterogeneous fields, our regular grid structure allows for a simpler velocity reconstruction that preserves the essential flow features. In this case, the longitudinal and transverse components of the velocity are obtained by arithmetic averaging due to the regular partitioning of the domain:
\begin{align}
\label{eq24}
u_1(\vx^{(j)}) =\frac{u_E^{(j)}+u_W^{(j)}}{2}, &&
    u_2(\vx^{(j)})=\frac{u_N^{(j)}+u_S^{(j)}}{2}, 
\end{align}
where $u_k^{(j)}$, with $k\in\sigma(j)$, are the edge velocities.
    
From the numerical flow simulations, we estimate the Eulerian velocity distribution $f_e(u)$ as
\begin{align}
\label{eq:fe2}
f_e(u_k) = \frac{1}{N_\mathrm{c}}
\sum_{j=1}^{N_\mathrm{c}}
\frac{\mathbb{I}\!\left[u_k \leq u_e(\mathbf{x}^{(j)}) < u_k + \delta u_k\right]}{\delta u_k},
\end{align}
where the sum runs over all $N_\mathrm{c}$ computational cells (or grid nodes), $u_k$ denotes the lower edge of the $k$-th velocity bin of width $\delta u_k$, and $\mathbb{I}[\cdot]$ is the indicator function.
Similarly, the advective tortuosity is estimated as
\begin{align}
\label{eq:tortuosity2}
\chi = \frac{\sum_{j} u_e(\vx_j)}{\sum_j u_1(\vx_j)}. 
\end{align}

\subsection{Hydrodynamic Transport in Fractures}
\label{sec:transport}
\subsubsection{Transport Model}
We consider purely advective transport in the rough fracture. Thus, the evolution of a scalar field $c(\vx,t)$ is described by the advection equation:
\begin{align}
\label{eq:advection}
\frac{\partial c(\vx,t)}{\partial t} = - \boldsymbol\nabla \cdot \left [ \vu(\vx)  c(\vx,t) \right ]. 
\end{align}

However, instead of solving this partial differential equation, we adopt a Lagrangian approach based on a time-domain random walk (TDRW) framework, in which fluid particle trajectories and residence times are explicitly tracked. The concentration field $c(\mathbf{x},t)$ is obtained a posteriori by ensemble averaging over the $10^7$ particle trajectories. The TDRW scheme is defined in such a manner that the $c(\mathbf{x},t)$ thus obtained coincides with the one that would be obtained by solving the advection equation \eqref{eq:advection} in the Eulerian framework. The fluid particles are thus infinitesimal, massless, and passive; they perfectly follow the local velocity field, without mechanical interaction with the fracture walls or each other. They are also appropriate for representing the transport of dilute conservative solute particles in the limit of very high Péclet numbers \citep{Zvikelsky2006,Dontsov2014,Cheng2025}.

\subsubsection{Time-Domain Random Walk\label{sec:tdrw}}

This equation is solved using a time-domain random walk (TDRW) scheme based on an upstream weighting scheme. In this scheme, fluid particles move between the control volumes of the regular grid used for the finite volume solution of the flow problem, according to
\begin{subequations}
\label{eq:tdrw}
\begin{align}
x_1^{(n+1)} = x_1^{(n)} + \xi^{(n)} \Delta x\frac{u_1(\vx^{(n)})}{|u_1(\vx^{(n)})|}, && x_2^{(n+1)} = x_2^{(n)} + (1 - \xi^{(n)}) \Delta x\frac{u_2(\vx^{(n)})}{|u_2(\vx^{(n)})|},
\end{align}
where $\xi^{(n)}$ is a random variable that takes the value 
1 with the transition probability toward the longitudinal downstream cell,
\begin{align}
\label{eq:wn}
w_{\parallel}^{(n)}= \frac{|u_1(\vx^{(n)})|}{|u_1(\vx^{(n)})| + |u_2(\vx^{(n)})|}  
\end{align}
and $0$ with probability $w_{\perp}^{(n)} = 1 - w_{\parallel}^{(n)}$, corresponding to movement into the transverse downstream cell. Note that $u_1(\vx^{(n)})$ and $u_2(\vx^{(n)})$ are the velocity components in the centres of the finite volumes. The transition time is given by
\begin{align}
t^{(n+1)} = t^{(n)} + \tau^{(n)}, && \tau^{(n)} = \frac{\Delta x}{|u_1(\vx^{(n)})| + |u_2(\vx^{(n)})|}. 
\end{align}
\end{subequations}
This expression reflects the total advective flux along the coordinate directions, rather than the Euclidean norm of the velocity, which would overestimate the effective displacement rate on a Cartesian grid. It ensures that the particle residence time is inversely proportional to the total outflow from the current control volume. In this way, residence times adapt to the local flow field: slower velocities yield longer dwell times, while faster velocities result in shorter steps. 
We provide more details about the equivalence between this scheme and the advection equation~\eqref{eq:advection} in Appendix~\ref{app:tdrw}. 

Note that, although the system is purely advective, stochastic routing is employed in this TDRW scheme to capture uncertainty in the direction of particle motion at the grid scale. The local velocity vectors are not necessarily aligned with the Cartesian grid, and the finite-volume discretization introduces ambiguity in determining the dominant direction of flow across cell faces. The stochastic routing scheme resolves this by assigning probabilistic transitions based on local velocity magnitudes, preserving mass conservation and compatibility with the underlying upwind discretization.  
Particle motion is restricted to transitions between adjacent control volumes to maintain consistency with the finite-volume discretization of the flow field. 
The TDRW scheme preserves compatibility with the fluxes computed at control volume interfaces, and ensures numerical stability. Moreover, the adopted routing scheme is equivalent to an upwind finite-volume advection solver and accurately captures the effects of flow heterogeneity. 
\begin{figure}
\centerline{\includegraphics[width=1.0\textwidth]{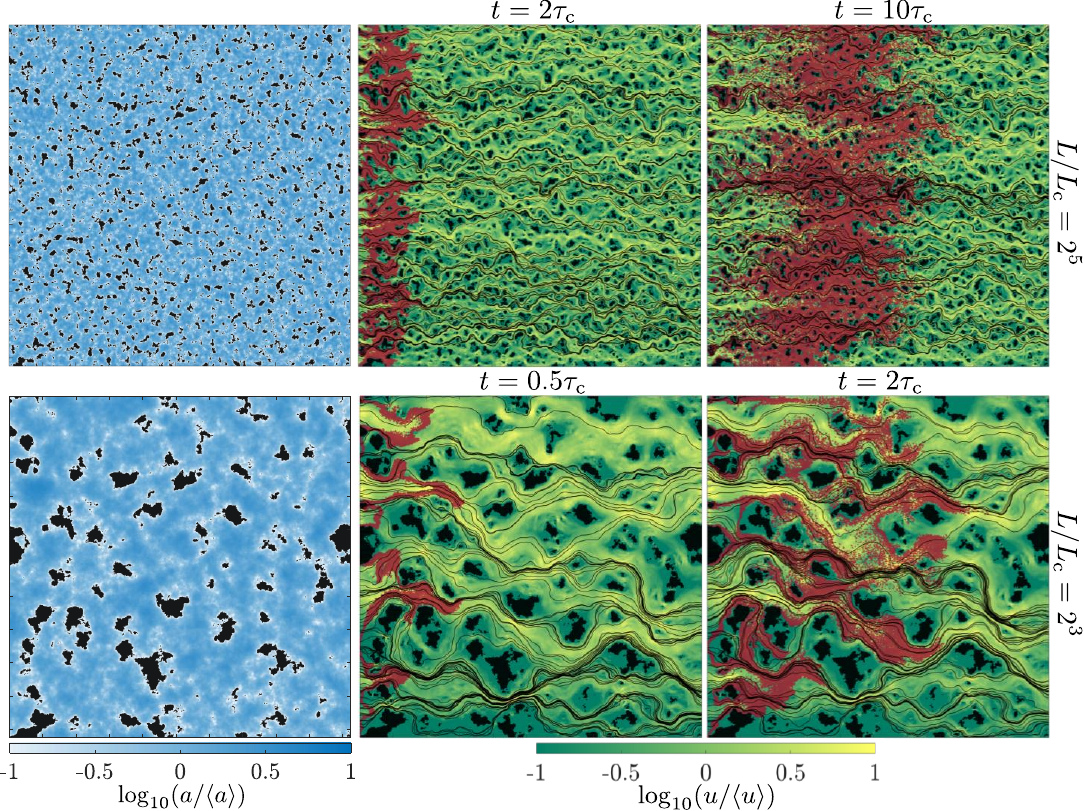}}
\caption{Maps of fracture apertures (left column) and the corresponding Eulerian velocity magnitude (middle and right columns) at two different times, with $10^7$ superimposed flux-weighted injected particles at the indicated times $t$, for two synthetic fractures with different correlation lengths, $L/L_\mathrm{c} = 2^3$ and $2^5$, for $\sigma_a/\langle a\rangle=0.75$, $L_\mathrm{c}=0.1~\textrm{m}$, and $\langle a\rangle=0.001~\textrm{m}$. Contact zones are depicted in black.
}
\label{Fig3}
\end{figure}
\subsubsection{Initial Particle Distributions}
The distribution of initial particle positions $\vx^{(0)}$ is denoted by $\rho(\vx)$. Particles are injected along a line on the inlet boundary at $x_1 = 0$, so that
\begin{align}
\rho(\vx) = \delta(x_1) \rho(x_2) .
\end{align}
The initial particle distribution and, thus, the initial velocity distribution, affects the average pre-asymptotic behavior, as demonstrated by several authors \citep{Hyman2015,Dentz2016,Fiori2017,Kang2017,Zech2018,Comolli2019,dentz2025linear}.

We employ two different initial particle distributions. First, we consider the uniform distribution, for which the particles are injected with a uniform density
\begin{align}
\label{eq:rho_uniform}
\rho(x_2) = \frac{1}{L_2},
\end{align}
where $L_2$ is the length of the initial line along the inlet boundary. Second, we consider a flux-weighted distribution, for which particles are injected with a density proportional to the velocity magnitude along the same line, that is,
\begin{align}
\label{eq:rho_fw}
\rho(x_2) = \frac{|\vu(0,x_2)|}{\displaystyle\int\limits_0^{L_2} d x_2 |\vu(0,x_2)|}.
\end{align}
In both injection modes, particles are released along a central portion of the inlet boundary, covering approximately half its length. This injection strategy avoids proximity to the lateral Neumann boundaries, ensuring that particle trajectories are not affected by boundary artifacts. Numerical tests confirmed that the results are insensitive to the injection location within this central region. Examples of particle plumes resulting from flux-weighted injection are presented in Fig.~\ref{Fig3} at different times. Here, $\tau_\mathrm{c}$ denotes the characteristic advection time, defined as the average time required for a fluid particle to travel a distance equal to the correlation length in the longitudinal direction.

\subsubsection{Observables}
In order to analyze particle transport through rough-walled fractures, for the two injection modes, we consider 
the mean and variance of streamwise displacement, as well as the distribution of first passage times through the fracture’s outlet control plane. 
The displacement mean and variance are defined analogously as
\begin{align}
\mathcal{M} (t) = \left \langle x_1^{(n_t)} \right \rangle, && \mathcal V(t) = \left \langle  \left (x_1^{(n_t)} \right )^2 \right \rangle - \left \langle x_1^{(n_t)} \right \rangle^2.
\end{align}
where the number of TDRW steps at time $t$ is given by
\begin{align}
n_t = \sup\{ n|t^{(n)} < t \}. 
\end{align}
The angular brackets denote the averaging over all particles.
The longitudinal dispersion coefficient, which describes the temporal
rate of change of the displacement variance, is given by
\begin{equation}
\mathcal{D}_\textrm{L}(t)=\frac{1}{2}\frac{d\mathcal{V}(t)}{dt}.
\end{equation}
The breakthrough times of particles at a control plane located at a distance $x_1$ from the inlet boundary along the mean flow direction is defined by
\begin{align}\label{eq:bt_time}
\mathcal{T}(x_1) = t^{(n_{x_1})}, \qquad 
n_{x_1} = \inf \left\{ n \mid x_1^{(n)} \ge x_1 \right\}.
\end{align}
The BTC or arrival time distribution at longitudinal position $x_1$ from the inlet boundary is defined by
\begin{equation}
\mathcal{F}(t,x_1) = \langle \delta[t-\mathcal{T}(x_1) ]\rangle.
\end{equation}

\subsection{Monte Carlo Simulations of Fracture Flow and Transport\label{sec:dns}}
In order to systematically study the impact of fracture heterogeneity on fracture scale flow and transport we perform a Monte-Carlo
analysis. The analysis is conducted by generating heterogeneous fracture aperture fields according to the four combinations of parameters listed in Table \ref{tab1}. For each combination, $N_\textrm{MC} = 100$ Monte Carlo (MC) realizations of the flow field are obtained by solving the Reynolds equation over a synthetic aperture field, partitioned in $2^{10}\times 2^{10}$ non-overlapping finite volumes. A plume of fluid particles is released along the inlet boundary using either uniform or flux-weighted injection, depending on the simulation scenario. Unless otherwise stated, figures and results referring to specific realizations use the flux-weighted mode, as noted in the captions. Table \ref{tab1} reports the combinations of parameters adopted to generate the synthetic aperture fields. In particular, we consider two relative closures, 0.25 and 0.75. The higher value (0.75) guarantees that the fractures exhibit a large flow heterogeneity while also ensuring that all fracture realizations maintain at least one connected flow pathway between the inlet and outlet boundaries and even that there is only one connected flow domain (i.e., no closed region isolates an open region from other open regions). This choice simplifies the Monte Carlo analysis by avoiding pathological cases. Note also that all results are nondimensionalized with respect to the mean aperture $\langle a \rangle$, which is set to $1~\mathrm{mm}$ in all simulations. The flow structure and transport behavior are fully characterized by the two dimensionless parameters $\sigma_a / \langle a \rangle$ and $L / L_\mathrm{c}$ (the Hurst exponent $H$ being always equal to 0.8).

\begin{figure}
\centering
\includegraphics[width=1\textwidth]{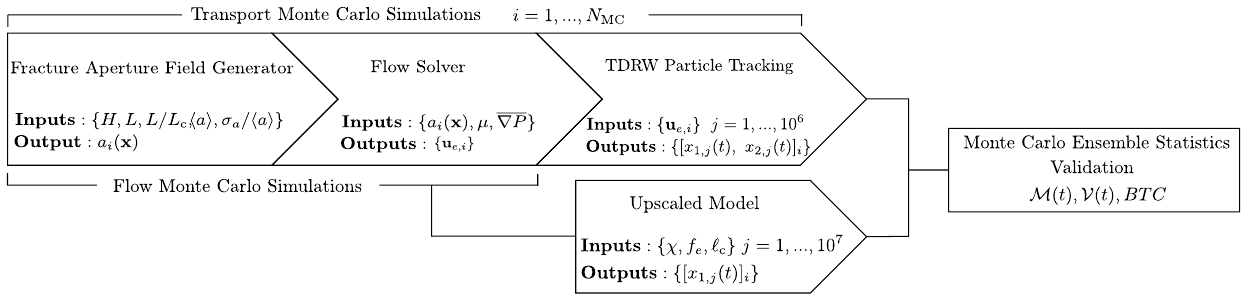}
\caption{Flow chart of the numerical modeling workflow. From geometry generation and flow simulation to particle transport and upscaling, all steps are embedded within the Monte Carlo framework.}
\label{Fig4}
\end{figure}

We also consider a correlation length $L_\mathrm{c} = 0.1~\mathrm{m}$, representative of the experimentally measured correlation lengths in fractured rock surfaces \citep{Brown1995,Meheust2000}, and vary $L/L_\mathrm{c}$ to assess the impact of fracture scale on advective transport. The synthetic aperture and velocity fields used in this study are grouped into four Monte Carlo datasets (MC1–MC4), each comprising 100 realizations. These datasets are publicly available via Zenodo~\citep{zenodo_run01, zenodo_run02, zenodo_run03, zenodo_run04}. After solving the flow problem for each realization and combination of parameters, the hydrodynamic transport is simulated using the TDRW scheme presented in Section \ref{sec:tdrw} and Appendix~\ref{app:tdrw}. Ensemble statistics are produced to analyze the average behavior across different realizations. 
\begin{table}\centering
\begin{tabular}{ccccccccccccc}
  \multicolumn{2}{c}{Monte Carlo}       & \multicolumn{5}{c}{Aperture Field Generator Inputs} & Fluid Properties&   \multicolumn{3}{c}{Upscaled Model Inputs} & Zenodo \\
 ID &  $N_\textrm{MC}$ & $L/L_\mathrm{c}$ & $\sigma_a/\langle a\rangle$ & $\langle a\rangle$                  & $L_\mathrm{c}$ &  $H$    & $\mu$ & $\ell_\mathrm{c}$ & $\chi$  & No. Part. & Dataset   \\ \hline
1       &  \multirow{4}{*}{100} & $2^5$            & 0.75                  & \multirow{4}{*}{$1~\textrm{mm}$} & \multirow{4}{*}{$0.1~\textrm{m}$} &  \multirow{4}{*}{0.8}  &\multirow{4}{*}{$10^{-3}~\mathrm{Pa\,s}$} &0.04~\textrm{m}& 1.12 & \multirow{4}{*}{$10^7$}& Run01\\ 
2       && $2^5$            & 0.25                  &                                    &    & &   &0.05~\textrm{m}&  1.01 &     & Run02 \\ 
3       && $2^3$            & 0.75                  &          & &    &  &0.05~\textrm{m}& 1.11  &     & Run03\\ 
4       && $2^3$            & 0.25                  &         &  &   & &0.04~\textrm{m}&  1.01  &     & Run04\\ 
\end{tabular}
\caption{Monte Carlo simulation sets (MC1–MC4) and corresponding dataset parameters. The associated Zenodo datasets (Run 01–04) are detailed in the Data Availability Statement.}\label{tab1}
\end{table}
The schematic flow chart (see Figure~\ref{Fig4}) summarizes the overall workflow of the study. The chart highlights the generation of synthetic aperture fields, the computation of flow using the Reynolds equation, the particle transport simulations via time-domain random walks (TDRW), and the application of the upscaled CTRW model. The entire pipeline is embedded in the Monte Carlo simulation scheme used to generate ensemble statistics across fracture realizations.

\subsection {Upscaled Transport Model}\label{upscale_model}
The TDRW model requires detailed knowledge of the full velocity field and resolved individual trajectories across the grid, which is computationally intensive for large domains or ensembles. We propose below an upscaled approach based on a correlated continuous time random walk (CTRW) that formulates transport as a one-dimensional Ornstein–Uhlenbeck process, replacing local dynamics with a stochastic description based on a few global parameters (e.g., velocity PDF, tortuosity, correlation length). This reduces computational costs while retaining the essential features of non-Fickian dispersion, making it well-suited for uncertainty quantification and predictive modeling across realizations. 
This approach has been used to quantify stochastic particle motion in heterogeneous flow fields at the pore scale \citep{Morales2017,Puyguiraud2019b} and Darcy scale~\citep{Comolli2019}, as well as at the fracture network scale \citep{Dentz2023}. 
We summarize its main elements below. 

We describe particle motion along a streamline parameterized by the curvilinear coordinate $s$. Its projection onto the mean flow direction is denoted by $x_1$. The coupled kinematics are
\begin{align}\label{eq:particle_motion}
\frac{dx_1(s)}{ds} = \frac{1}{\chi}, \qquad \frac{dt(s)}{ds} = \frac{1}{u_s(s)},
\end{align}
where $\chi$ is defined by Eq.~\eqref{eq:tortuosity2} and $u_s(s)$ is the Lagrangian particle velocity along the streamline. The particle velocities $u_s(s)$ are distributed according to the flux-weighted Eulerian velocity PDF, which is defined from the Eulerian velocity PDF $f_e$ as \citep{Dentz2016},  
\begin{align}\label{eq:pdf_relationship}
f_s(u)=\frac{u f_e(u)}{\langle u_e\rangle} .
\end{align}
The initial velocity distribution $f_0(u)$ is
\begin{align}
\label{eq:f0u}
f_0(u) = f_e(u)
\end{align}
for the uniform injection mode and
\begin{align}
\label{eq:f0fw}
f_0(u) = f_s(u) 
\end{align}
for the flux-weighted injection mode. Note that this equivalence requires ergodicity of the initial line in the sense that along the initial line a representative part of the
velocity statistics can be sampled. 

The series of Lagrangian velocities $\{u_s(s)\}$ is modeled as a stationary Markov process \citep{Dentz2016,Hakoun2019} characterized by the transition probability $r(u,s|u')$ and the stationary distribution $f_s(u)$. This implies that we assume Lagrangian ergodicity, that is, that a particle is able to sample the full velocity statistics along a sufficiently long streamline. In this framework, the distribution of particles along the mean flow direction is given by
\begin{align}
c(x_1,t)=\left\langle\delta\left[x_1 - s(t) / \chi\right]\right\rangle ~  && \mathrm{with} ~ ~ ~ s(t) = \sup(s|t(s) \leq t) , 
\end{align}
where $\langle \cdot\rangle$ represents the average on all particles. The displacement mean and variance are, respectively,
\begin{align}
\mathcal{M} (t )=\frac{\langle s(t)\rangle}{\chi} \text{~ ~ and ~ ~} \mathcal{V}(t)=\frac{\langle[s(t)-\langle s(t)\rangle]^2\rangle}{\chi^2}. 
\end{align}
The BTC is obtained from the particle times $t(s)$ as
\begin{equation}
\mathcal{F}(t,x_1) = \langle\delta[t-t(x_1/\chi)]\rangle.
\end{equation}
It can be shown that the joint distribution of longitudinal particle position and velocity $f(x_1,u,t) = \langle\delta[x_1 - x_1(t)]\delta[u-u(t)]\rangle$ satisfies the integro-differential equation \citep{Comolli2019}
\begin{equation}
\frac{\partial f(x_1,u,t)}{\partial t}+\frac{u}{\chi}\frac{\partial f(x_1,u,t)}{\partial x_1}=-\frac{u}{\Delta s}f(x_1,u,t)+\int\limits_0^\infty du^\prime \frac{u^\prime r(u,\Delta s|u^\prime)}{\Delta s}f(x_1,u^\prime,t). 
\end{equation}
The second term on the left side of the equation describes the translation of the distribution by the local velocity, the first term on the right side transitions away from the current velocity, and the second toward the current velocity. 

The evolution of $u_s$ is modeled through an Ornstein-Uhlenbeck process for the normal score transform $z(s)$ of $u_s(s)$, which is defined by
\begin{align}
\label{eq:us}
z(s)=\Phi^{-1}\{F_s[u_s(s)]\} \text{~ or, equivalently, ~} u_s(s)=F_s^{-1}\{\Phi[z(s)]\},
\end{align}
where $F_s(u)$ is the cumulative distribution function (CDF) of the $s$-Lagrangian velocity and $\Phi(z)$ is the CDF of the Gaussian distribution of zero mean and unit variance. The evolution of $z(s)$ is then defined by the Ornstein-Uhlenbeck process \citep{Gardiner2009}
\begin{equation}\label{eq:Langevin_equation2}
\frac{dz(s)}{ds} = -\frac{z(s)}{\ell_\mathrm{c}}+\sqrt{\frac{2}{\ell_\mathrm{c}}}\eta(s),
\end{equation}
where $\eta(s)$ is a Gaussian white noise with zero mean and correlation $\langle \eta(s)\eta(s') \rangle = \delta(s - s')$. The parameter $\ell_\mathrm{c}$ sets the characteristic correlation length of $u_s(s)$. As shown by \citet{Lenci2024}, the correlation length of the flow field increases more rapidly than that of the underlying aperture due to channel connectivity. These findings support interpreting the dynamic correlation scale $\ell_\mathrm{c}$ introduced here as an emergent Lagrangian persistence length, typically smaller than the nominal geometric correlation length $L_\mathrm{c}$.

The numerical implementation of this CTRW approach is based on the discretized versions of Eq.~\eqref{eq:particle_motion},
\begin{align}
x_1(s + \Delta s) = x_1(s) + \frac{\Delta s}{\chi}, && t(s + \Delta s) = t(s) + \frac{\Delta s}{u_s(s)},
\end{align}
and~\eqref{eq:Langevin_equation2},
\begin{align}
z(s+\Delta s) = z(s)\left(1-\frac{\Delta s}{\ell_\mathrm{c}}\right)+\sqrt{2\frac{\Delta s}{\ell_\mathrm{c}}}\eta(s),
\end{align}
where $u_s(s)$ is given by the expression~\eqref{eq:us}. 

To be applicable to a single fracture realization, this framework must assume Lagrangian ergodicity, meaning that particle trajectories sample the full velocity distribution along sufficiently long streamlines. This assumption may be violated in systems with limited spatial extent or large correlation lengths. In such cases, statistical fluctuations and structural constraints reduce the representativeness of the local velocity field, potentially affecting the predictive accuracy of the upscaled model. 

However, in this study, the CTRW framework is adopted to analyze the pre-asymptotic transport behaviors and the evolution of the following ensemble-averaged quantities (i.e., averaged over the fracture population) obtained from the detailed numerical simulations described in Section \ref{sec:dns}: mean longitudinal displacement, mean longitudinal variance, as well as mean BTC at the fracture's outlet. Hence, flow ergodicity in individual fractures is not required for the upscaled model to accurately predict the results obtained with direct (TDRW-based) simulations. The upscaled model therefore requires the following inputs: (i) the Eulerian velocity PDF ($f_e(u)$), (ii) the advective tortuosity ($\chi$), and (iii) the characteristic length of the flow ($\ell_{\textrm{c}}$). These quantities are obtained by averaging over the Monte Carlo flow simulations. The correlation length $\ell_\mathrm{c}$ is estimated by fitting the upscaled model to the late-time evolution of the breakthrough curves obtained from the TDRW simulations. This approach allows us to identify the correlation length that best reproduces the ensemble transport dynamics observed in the full-resolution model. The parameters for the upscaled CTRW model are reported in Table \ref{tab1}.

\section{Fracture Scale Flow and Transport Behaviors}
\label{sec:results}
In this section we discuss the flow and transport behaviors in single rough fractures for the heterogeneity scenarios given in Table \ref{tab1}. We first analyze the velocity statistics in light of the upscaled CTRW model presented in the previous section. Then, we discuss the advective transport behaviors obtained from the detailed numerical flow and transport simulations and compare them to the behaviors predicted by the CTRW model.   

\subsection{Velocity Fields and Velocity Probability Density Functions}
Figure~\ref{Fig3} provides two examples of aperture fields generated with a relative closure of $\sigma_a/\langle a \rangle = 0.75$, and with $L/L_\mathrm{c} = 2^3$ and $2^5$. The corresponding stationary velocity fields are shown as well. The flow patterns vary on the correlation length $L_\mathrm{c}$; hence this length scale controls the spatial extent of flow channeling \citep{Meheust2003, Lenci2024}. The probability density function (PDF) of the Eulerian velocity offers insights into the impact of the medium's heterogeneity on that of the flow, enabling the characterization of transport properties based on flow statistics. Specifically, the behavior of the PDF at high velocities can be associated with preferential flow channels and governs the medium's transmissivity. In contrast, the low velocity behavior of the PDF is related to the occurrence of quasi-stagnant zones, which dominate the late time scaling of solute transport. 

In fact, the plume evolution is driven by velocity contrasts resulting from the medium's heterogeneity: the leading edge of the plume rapidly migrates through high-velocity channels, while the trailing edge remains trapped in quasi-stagnant zones. The mechanism of advective spreading dominates transport for times much longer than the characteristic diffusive time scale of the medium \citep{Andrade1997,Dentz2004,Tyukhova2016}. Therefore, understanding flow heterogeneity, particularly low-velocity behavior, constitutes a powerful tool for characterizing hydrodynamic transport.
\begin{figure}
\centerline{\includegraphics[width=0.8\textwidth]{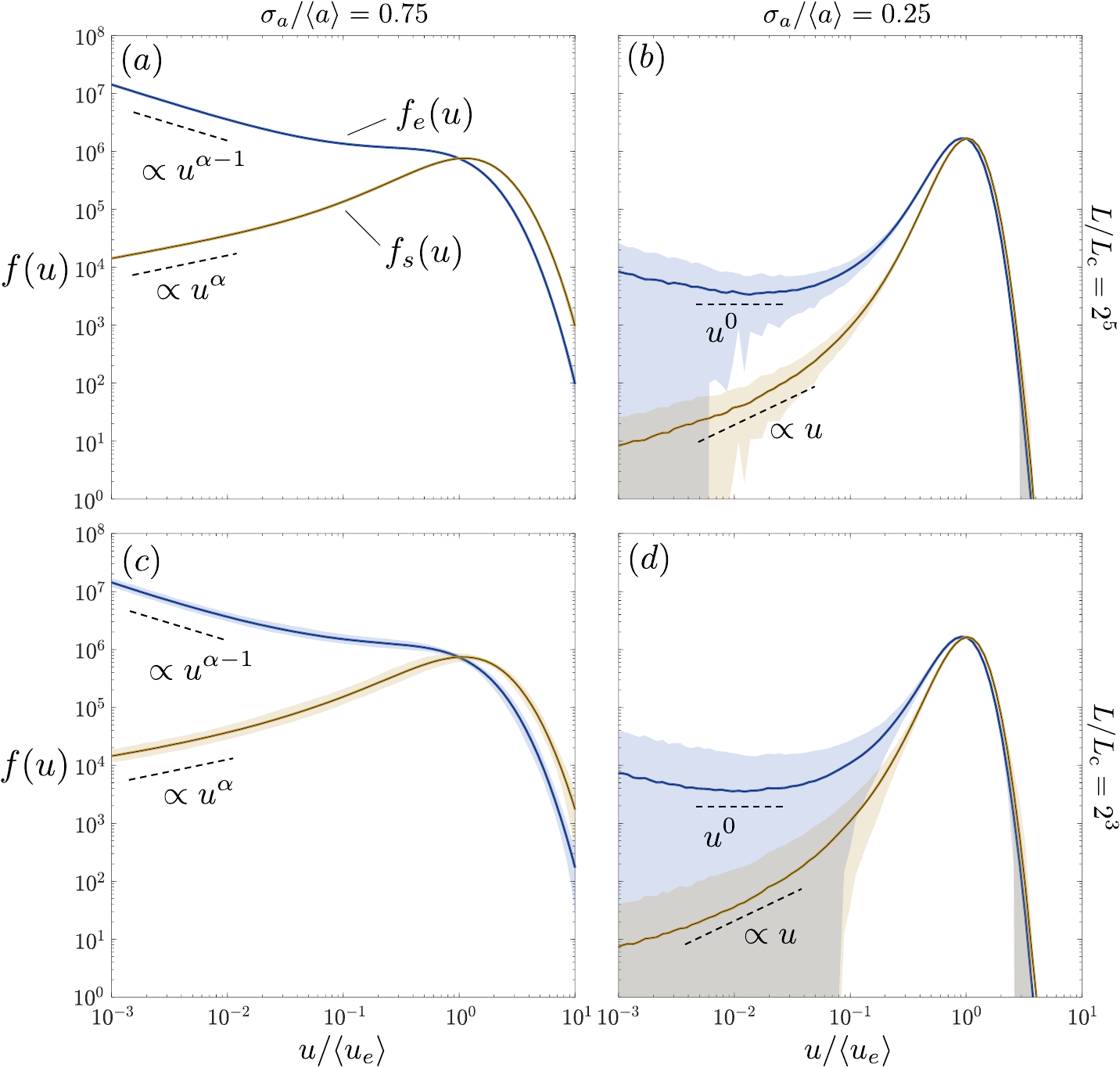}}
\caption{
Probability density functions (PDFs) of Eulerian (blue) and $s$-Lagrangian (yellow) velocities for the four Monte Carlo realizations: (a) MC1, (b) MC2, (c) MC3, and (d) MC4. The corresponding parameters used for aperture field generation are listed in Table~\ref{tab1}. Trend lines emphasize the scaling behavior of the low-velocity tails: the Eulerian PDF scales as $u^{\alpha - 1}$ (dashed line), while the $s$-Lagrangian PDF scales as $u^{\alpha}$ (dash-dotted line), in agreement with theoretical predictions for transport in heterogeneous flow fields. The shaded areas represent the confidence interval between the 5th and 95th percentiles.
}
\label{Fig5}
\end{figure}
Figure~\ref{Fig5} displays the normalized Eulerian velocity PDFs, used as inputs for the upscaled CTRW model, together with the corresponding $s$-Lagrangian PDFs obtained by flux weighting, as described in Eq.~\eqref{eq:pdf_relationship}. 
The PDFs for the same relative closures (Figures~\ref{Fig5}(a) and~\ref{Fig5}(c) for $\sigma_a/\langle a \rangle = 0.75$, and Figures~\ref{Fig5}(b) and~\ref{Fig5}(d) for $\sigma_a/\langle a \rangle = 0.25$) are virtually identical, whereas the relative correlation length plays only a minor, or no, role. Thus, the aperture heterogeneity clearly dominates the behavior of the velocity PDF, while the correlation length has virtually no impact on the flow velocities' ensemble statistics, at least for the values under consideration. This can be understood by the fact that the ratio between fracture length and correlation length is an estimate for the number of independent samples of aperture values, for example within a single realization. Thus, the ratio would affect the flow velocity statistics determined for a single aperture field. Here, however, we are considering ensemble statistics, which are obtained by sampling across different realizations.

These behaviors are proper to flow in random media, be it at the pore, continuum or regional scales, as long as a characteristic heterogeneity length scale exists such that the media can be considered ergodic. Ergodicity means that the statistics sampled in a single medium realization are representative of the ensemble statistics. For the more heterogeneous aperture distribution of $\sigma_a/\langle a \rangle = 0.75$, we observe a high frequency of low velocities expressed by the power-law behavior $f_e(u) \propto u^{\alpha - 1}$ with $\alpha = 0.40$ and $f_s(u) \propto u^{\alpha}$ correspondingly. For $\sigma_a/\langle a \rangle = 0.25$, $f_e(u)$ has a peak at around the mean velocity and then decreases for decreasing velocity towards a plateau at small values. In the limit $\sigma_a/\langle a \rangle = 0$ (parallel plate fracture), the velocity PDF is expected to approach a delta function corresponding to uniform flow; that is, the smaller the relative closure, the higher and narrower the peak becomes. At low velocity a uniform asymptote $f_e(u) \propto u^0$ is observed, and $f_s(u) \propto u$ correspondingly. Similar behaviors for the velocity PDFs in media of different heterogeneity have been observed for flow in porous media at the pore and continuum scales \citep{deAnna2017, Hakoun2019, Souzy2020}.

\subsection{Displacement Statistics}
\subsubsection{Mean Displacement}\label{sec:meanDisplacement}
At times much shorter than the characteristic advection time, $\tau_\textrm{c}=\ell_\mathrm{c}/\langle u_1\rangle$, the particle velocities are approximately constant and equal to the initial velocities. Thus, the mean displacement is given by 
\begin{equation}
\mathcal{M}(t)=\int  d x_2 \rho(x_2) u_1(0,x_2) t \quad \textrm{for} \quad t\ll \tau_\textrm{c}. 
\end{equation}
For $t \ll \tau_\textrm{c}$, the CTRW model thus estimates 
\begin{equation}
\mathcal{M}(t) = \frac{1}{\chi} \int_0^\infty du f_0(u) u t. 
\end{equation}
For uniform injection, $\rho(x_2)$ is given by Eq. \eqref{eq:rho_uniform}. Under ergodic conditions, the initial velocity distribution is $f_0(u) = f_e(u)$ and thus both the direct simulations and CTRW model give
\begin{align}
\mathcal M(t) = \langle u_1 \rangle t.
\end{align}
For the flux-weighted injection, $\rho(x_2)$ is given by Eq.~\eqref{eq:rho_fw}. Under ergodic conditions, the corresponding initial velocity distribution is $f_0(u) = f_s(u)$. Thus, the mean displacement is 
\begin{align}
\mathcal M(t) = \langle u_1 \rangle \frac{\langle u_e^2 \rangle}{\langle u_e \rangle} t. 
\end{align}
For times $t \gg \tau_\textrm{c}$, the time evolution of the centre of mass is
\begin{align}
\mathcal M(t) = \langle u_1 \rangle t
\end{align}
because the stationary velocity distribution is given by the Eulerian flow statistics. Thus, for an ergodic source, the mean displacement for the uniform distribution should be stationary and given by $\langle u_1 \rangle t$. For the flux-weighted injection, the mean particle velocity decreases from the higher flux-weighted mean towards the Eulerian mean velocity. 

\begin{figure}
\centerline{\includegraphics[width=0.8\textwidth]{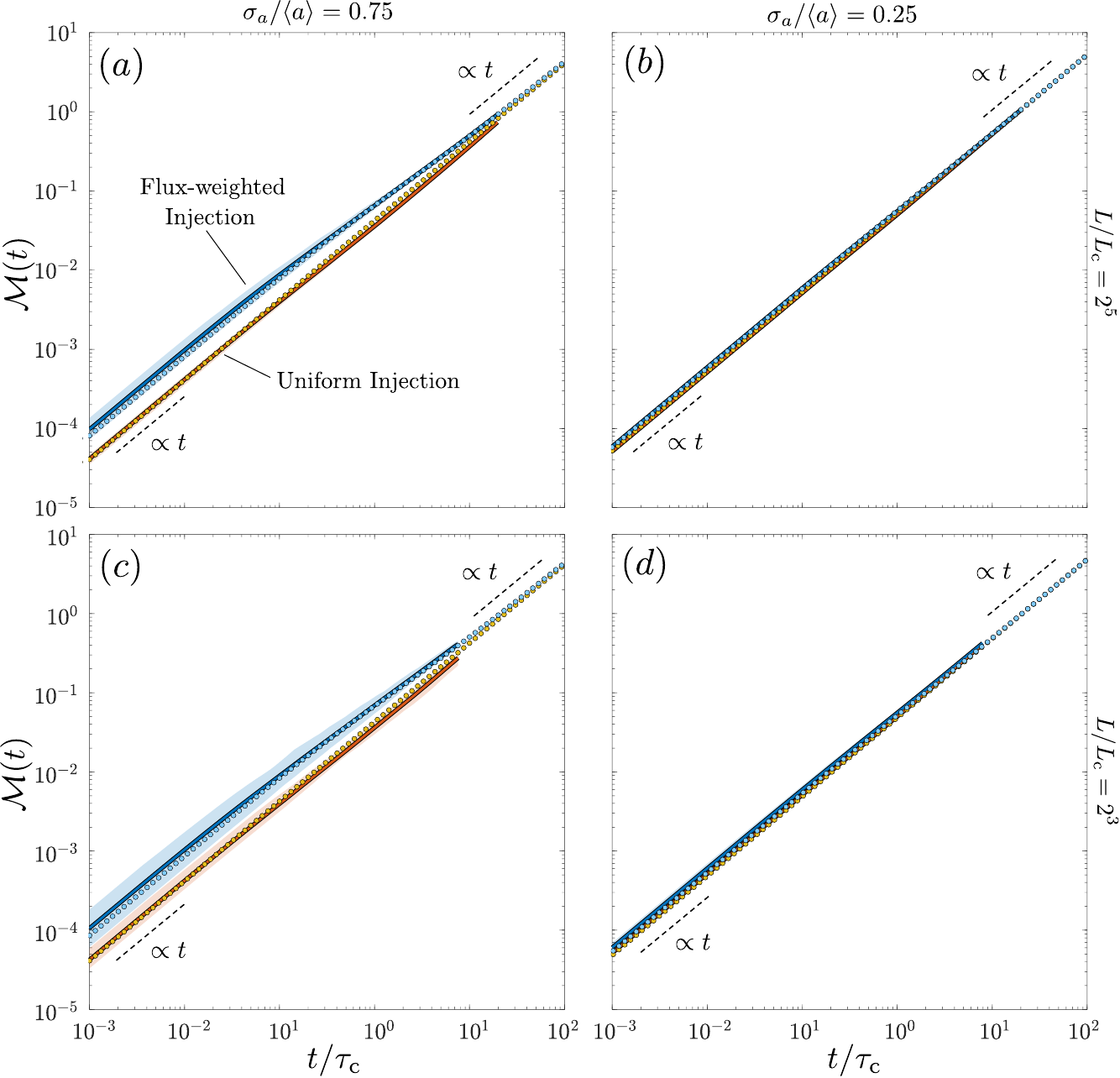}}
\caption{Mean displacement for uniform (orange) and flux-weighted (blue) injection, obtained from direct simulations (solid lines) and the upscaled model (symbols). The fracture aperture field parameters used in each case are reported in Table \ref{tab1} for the four parameter combinations: (a) MC1, (b) MC2, (c) MC3, and (d) MC4. Shaded areas represent the confidence interval between the 5th and 95th percentiles.}
\label{Fig6}
\end{figure}

Figure~\ref{Fig6} shows the evolution of the mean displacement for both uniform and flux-weighted distributions from the detailed Monte-Carlo simulations and the upscaled CTRW model for the scenarios given in Table \ref{tab1}, in good agreement with the behaviors expected from the theoretical CTRW predictions. The upscaled model captures the full evolution of the mean displacement.
In Figure~\ref{Fig6} the confidence intervals show the variability of the mean displacement in individual fracture realizations around the ensemble mean behavior, between the 5th and 95th percentiles.  
While the ensemble mean behavior is of course independent of the fracture size, the variance is larger for the small fractures. The variability between fracture realizations decreases with increasing fracture size because large fractures are more representative of the ensemble statistics than small fractures. 
Hence, the only difference between Figures~\ref{Fig6}(a) and~\ref{Fig6}(c) (as well as between~\ref{Fig6}(b) and~\ref{Fig6}(d)) lies in the fact that the confidence intervals are consistently larger for the shorter fractures than for the longer ones.

\subsubsection{Displacement Variance}
At times $t \ll \tau_\textrm{c}$, the displacement variance increases ballistically, that is, 
\begin{equation}\label{eq:ballistic_variance}
\mathcal{V}(t) = \sigma_{u_{1,0}}^2 t^2, 
\end{equation}
where $\sigma_{u_{1,0}}^2$ is the variance of the initial particle velocity. The CTRW model approximates this short time behavior by
\begin{equation}\label{eq:prediction_variance}
\mathcal{V}(t) = \frac{\sigma_{u_e}^2}{\chi^2} t^2, 
\end{equation}
which slightly underestimates the true early time evolution~\citep{Comolli2019}.

For times $t \gg \tau_\textrm{c}$, the CTRW approach predicts superdiffusive behavior for the more heterogeneous fracture with $\sigma_a/\langle a \rangle = 0.75$ as \citep{shlesinger1974}
\begin{align}
\mathcal V(t) \propto t^{2 - \alpha}.
\end{align}
The exponent $\alpha$ characterizes the behavior of $f_e(u) \propto u^{\alpha -1}$ for $u \ll \langle u_e \rangle$, for which we find $\alpha = 0.4$ (see Figure~\ref{Fig5}(a,c)).

For the less heterogeneous fractures with $\sigma_a/\langle a \rangle = 0.25$, we find $\alpha = 1$ (see Figure~\ref{Fig5}b,d). In this case, the CTRW approach predicts \citep{shlesinger1974}
\begin{align}
\mathcal V(t) \propto t \ln t,
\end{align}
see also Appendix~\ref{app:scalings}. 

\begin{figure}
\centerline{\includegraphics[width=0.8\textwidth]{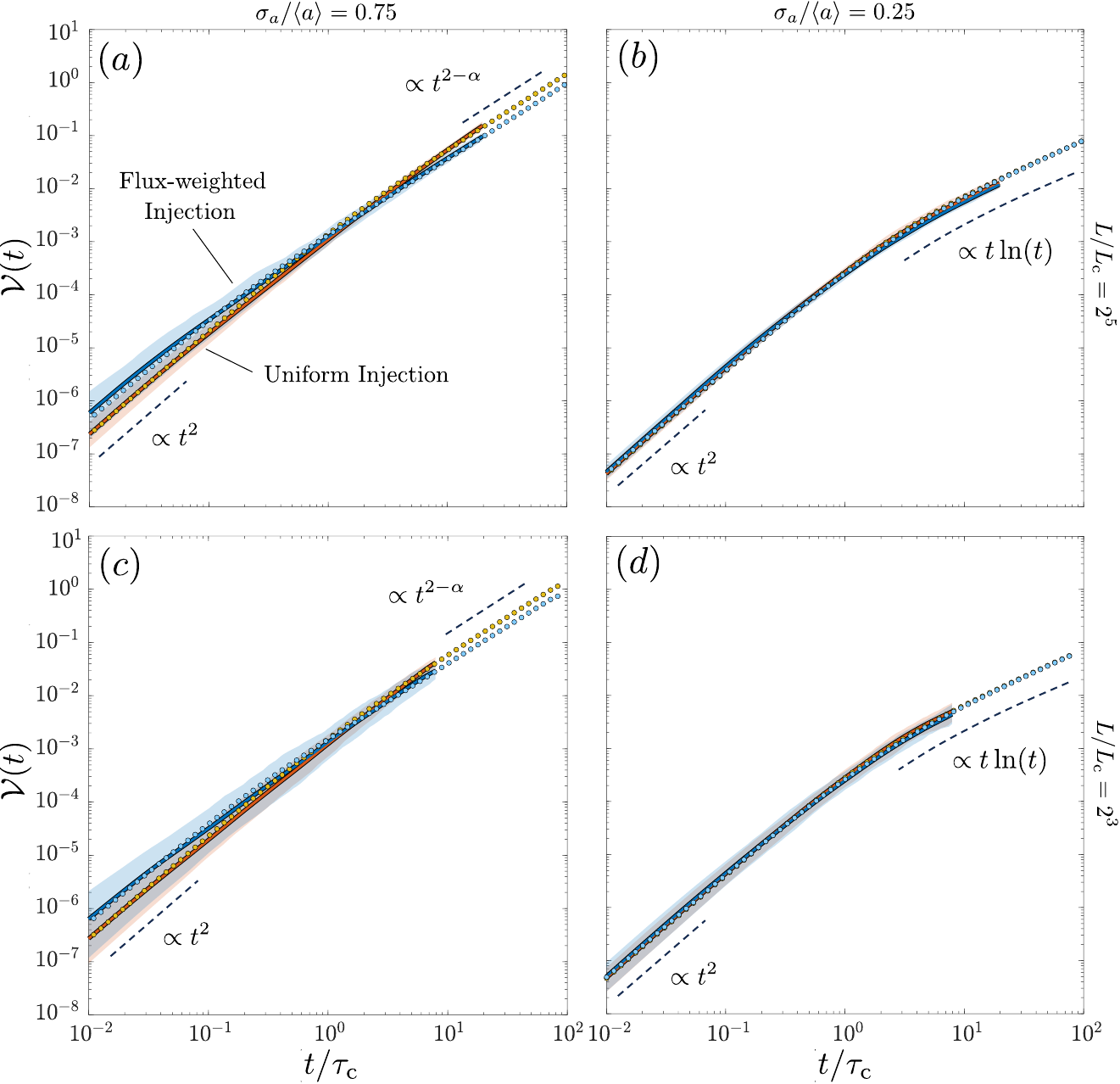}}
\caption{Displacement variances for uniform (orange) and flux-weighted (blue) distribution obtained from direct simulations (solid lines) and the upscaled model (symbols). Parameters used for fracture aperture fields generation are reported in Table \ref{tab1}: (a) MC1,  (b) MC2,  (c) MC3, and  (d) MC4. Shaded areas represent the confidence interval between the 5th and 95th percentiles.}
\label{Fig7}
\end{figure}

Figure~\ref{Fig7} displays the evolution of the displacement variance from the detailed Monte-Carlo simulations and the upscaled CTRW model. The data feature the ballistic early time behavior and the cross-over to the superdiffusive scaling predicted by the CTRW model based on the Eulerian velocity statistics, for both relative fracture closures. 
The upscaled CTRW model provides an excellent match with the detailed numerical simulations. The ensemble mean behavior is of course the same for the large and small fractures. As discussed above for the mean displacement (section~\ref{sec:meanDisplacement}), the variability around the ensemble mean, indicated by the shaded regions in Figure~\ref{Fig7}, decreases with increasing fracture size.  

\subsection{Breakthrough Curves}
\begin{figure}
\centerline{\includegraphics[width=0.8\textwidth]{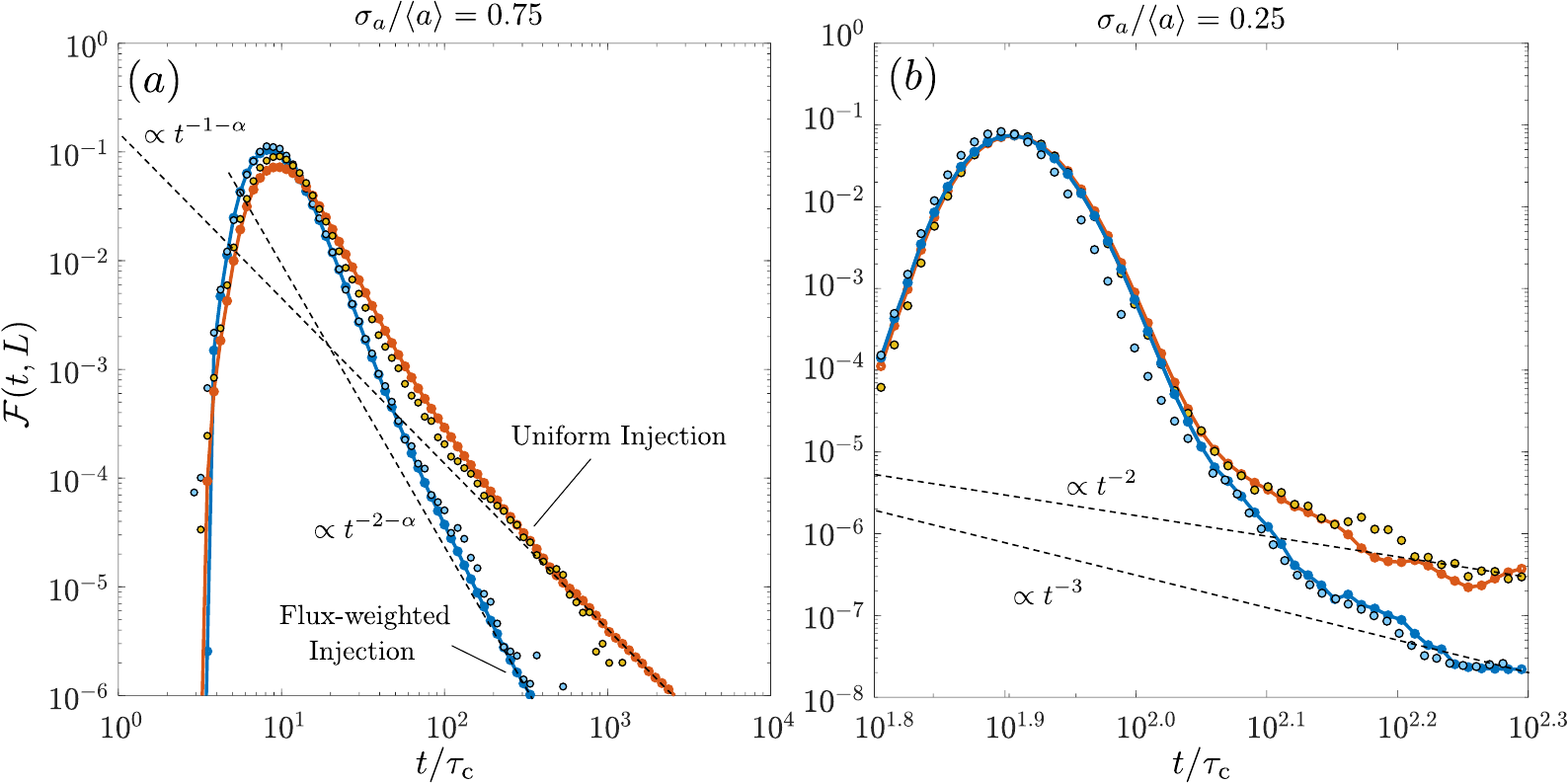}}
\caption{Breakthrough curves for uniform (orange) and flux-weighted (blue) injections, obtained from direct simulations (lines with dark-colored dots) and the upscaled model (light-colored dots). BTCs are evaluated at the fracture outlet, located at a longitudinal distance $L$ from the injection front. Synthetic fractures were generated with $L_{\mathrm{c}} = 0.1\,\mathrm{m}$ ($L/L_\mathrm{c}=2^5$) and average aperture $\langle a\rangle = 0.001\,\mathrm{m}$. The parameters used to generate the fracture aperture fields are listed in Table~\ref{tab1}. Only MC1 and MC2 are shown, as MC3 and MC4 exhibit similar behavior due to averaging over a sufficiently large Monte Carlo ensemble. Note the different tail exponents dictated by $\alpha$: $\alpha=0.4$ for $\sigma_a/\langle a\rangle=0.75$ and $\alpha=1$ for $\sigma_a/\langle a\rangle=0.25$.}
\label{Fig8}
\end{figure}
Figure~\ref{Fig8} shows BTC data for the large fracture ($L/L_\mathrm{c}=2^5$) at the two investigated relative fracture closures, obtained both from the particle tracking simulations and from the upscaled CTRW model. The graphs for the smaller fractures ($L/L_\mathrm{c}=2^3$) with the same fracture closures are not shown, but, as for the other observables (plume position and variance), we would expect the same mean behavior for the smaller fractures as for the larger fractures.
The upscaled (CTRW) model provides a good match with the features of the BTCs inferred from the direct simulations, including their late time power law behavior, both for uniform and flux-weighted initial conditions, and for both fracture closures.

The BTCs depicted in Figure~\ref{Fig8}(a) refer to the parameter combination MC1 (ratio $L/L_\mathrm{c}=2^5$ and relative closure $\sigma_a/\langle a\rangle=0.75$). Power law tails in the form $\mathcal F(t,L) \propto t^{-\gamma}$ are obtained at large times, with exponents of absolute values $\gamma = 1.4$ and $2.4$ for the uniform and the flux-weighted injection, respectively. 
These scalings are consistent with the power-law scaling observed in the velocity PDFs shown in Figure~\ref{Fig5} at low flow velocities. Indeed, the CTRW approach relates the power exponent of the velocity PDF to the BTC scaling \citep{Comolli2019, Dentz2023}. For $f_s(u) \propto u^{\alpha}$ and flux-weighted initial velocity distribution, it predicts that $\mathcal F(t,L) \propto t^{-2-\alpha}$ (here, $\alpha = 0.4$). For a uniform initial distribution, the BTC tailing is dominated by the long-waiting times at the origin. It thus scales as $\mathcal F(t,L) \propto t^{-1-\alpha}$ 
because the initial velocity PDF scales as $f_0(u) \propto u^{\alpha-1}$. 

Figure~\ref{Fig8}(b) shows results for simulations obtained  for $L/L_\mathrm{c}=2^5$ and relative closure $\sigma_a/\langle a\rangle=0.25$. This case is characterized by aperture fields of smaller variability with respect to the previous case, but the same ratio of the fractures' length to their correlation length. As expected, the peak of the BTC is much narrower than that obtained for a closure of 0.75 (see Figure~\ref{Fig8}(b)); in the limit of zero relative closure, it would approach a delta function. The velocity PDF scales as $f_s(u) \propto u$ at low velocities and the initial velocity as $f_0(u) \propto $ constant (i.e., the same scalings as for the relative closure 0.75, but with $\alpha=1$ instead of $\alpha=0.4$). Thus we expect the BTC tailing of $\mathcal F(t,L) \propto t^{-2}$ for the uniform and $\mathcal F(t,L) \propto t^{-3}$ for flux-weighted injection. The detailed numerical simulations confirm these scalings.

Note that, at sufficiently large relative closure $\sigma_a/\langle a\rangle$, since the aperture PDF remains finite at $a=0$ due to contact zones, a change of variables from $a$ to $T$ yields a power law asymptotic low-transmissivity behavior for the PDF of local transmissivities:
\[
f_T(T) \propto f_a(a)\left|\frac{da}{dT}\right|
\sim_{T\to 0} f_a(0) T^{-2/3}~.
\]
Using an approach where local transmissivities were prescribed as a log-normal distribution, i.e., $Y=\ln T$ prescribed as a Gaussian random
field with the variance $\sigma_Y^2$ and correlation length as independent
statistical parameters, \citet{Fiori2015} identified the range $[-1;0]$ of exponents for the low-$T$ behavior of $f_T(T)$ as that for which persistent anomalous transport arise. The present value $-2/3$ falls into that range, which explains that a similar qualitative transport behavior is seen here, with a BTC long time power law exponent also smaller than -2; however, in contrast to the absolute values of up to 6 obtained by these authors for large heterogeneities, here the exponent does not go below -3. As expected, since the BTC long times power law scaling is controlled by the low velocity behavior of the velocity PDF, it depends on the details of the local transmissivity PDF and correlation structure.

\section{Summary and Conclusions}
\label{sec:conclusions}
In this work, flow properties and hydrodynamic transport in realistic synthetic rough fracture geometries have been analysed from random walk simulations and an upscaled CTRW-based model, using a Monte Carlo (MC) framework to account for the stochasticity of the fracture realizations.  
To the best of our knowledge, this is the first time that such an approach has been applied to systematically characterize the anomalous features of hydrodynamic transport in single geological fractures. 

One hundred independent realistic synthetic rough fracture aperture fields were generated numerically with a prescribed relative closure of $\sigma_a / \langle a \rangle = 0.75$ and a prescribed correlation length $L_\mathrm{c}$, below which the aperture fields are self-affine. For each of them, another aperture field with the same fluctuations but a mean value such that $\sigma_a / \langle a \rangle = 0.25$, was also obtained by subtracting the appropriate constant value. This procedure was performed for two different fracture sizes, such that the ratio of the fracture length to the correlation length was either $2^3$ or $2^5$. Thus, one hundred fracture realizations were obtained for each of the four different sets of statistical geometrical parameters defining the fracture geometry, for the two $L/L_\mathrm{c}$ ratios and the two relative closures. 
A finite-volume scheme was employed to solve the steady, isothermal, depth-averaged flow in all heterogeneous aperture fields under the lubrication approximation. Purely advective transport was then simulated within each resulting velocity field using two complementary approaches: (i) a time-domain random walk (TDRW) particle-tracking scheme based on an upstream-weighting formulation, and (ii) a one-dimensional continuous-time random walk (CTRW) upscaled model. 
The TDRW scheme is well suited for parallel computation and computationally efficient, enabling the tracking of up to $10^7$ particles across multiple fracture aperture realizations. Both the TDRW and CTRW models were run for two injection modes: uniform and flux-weighted.

The results were analyzed in terms of five quantities of interest: the Eulerian and Lagrangian velocity PDF, the time evolution of the mean plume position, that of the plume's longitudinal variance, and the breakthrough curves (BTCs) at the fracture's outlet. For each of them, we considered the mean behavior over the population of one hundred fractures. We also considered a confidence interval around that mean behavior, defined by the 5th and 95th percentiles (except for the BTCs).

The velocity PDFs (either Eulerian or Lagrangian) do not depend on the correlation length $L_\mathrm{c}$, because they are sampled across all realizations of the aperture field.  
However, statistical fluctuations 
between velocity PDFs sampled in individual realizations are all the larger as the ratio $L/L_\mathrm{c}$ is smaller, because $L/L_\mathrm{c}$ controls the number of statistically independent velocity values within a single realization. For the same reason, the other quantities of interest exhibit a similar dependence on the correlation length: their ensemble average does not depend on it, while the fluctuations around that mean for a finite number of realizations in the fracture population are all the larger as the ratio $L/L_\mathrm{c}$ is smaller. In contrast, the closure of the fracture strongly impacts the velocity PDF: the more closed the fracture is, the more the PDF deviates from a peaked shape; the Eulerian PDF, in particular, decreases monotonically for a relative closure of 0.75 (for which $\sim 10 \% $ of the fracture plane is closed). The tails of the velocity PDFs exhibit a power-law behavior ($f(u) \propto u^{\alpha -1}$ for the Eulerian PDF, $f(u) \propto u^{\alpha}$ for the Lagrangian PDF) for low velocities, as expected from theoretical considerations.
The mean plume displacement exhibits a linear scaling as a function of time at both short and long times, and for both uniform and flux-weighted injections. The variance 
varies ballistically at short times (i.e., as the square of time), and exhibits a long time behavior controlled by the small velocity scaling of the velocity PDF, characterized by the exponent $\alpha$, which is 0.4 for the larger investigated closure ($\sigma_a/\langle a\rangle = 0.75$), and 1 for the smaller investigated closure ($\sigma_a/\langle a\rangle = 0.25$).
From CTRW theory, the large time scaling of the mean variance is then expected to be $t^{2-\alpha}$ and $t\, \ln t$ for $\sigma_a/\langle a\rangle = 0.75$ and $\sigma_a/\langle a\rangle = 0.25$, respectively. Both these long time scalings and the short time ballistic scaling are consistent with the behaviors obtained from the detailed numerical simulations.
 
The BTC peaks become narrower as the fracture closure decreases, and the scaling of their long-time power-law tails exhibits exponents in the form $-1-\alpha$ (for uniform injection) and $-2-\alpha$ (for flux-weighted injection), as expected from the CTRW theory, with $\alpha=0.4$ for $\sigma_a/\langle a\rangle = 0.75$ and $\alpha=1$ for $\sigma_a/\langle a\rangle = 0.25$.

The one-dimensional upscaled CTRW model, based on an Ornstein-Uhlenbeck stochastic process, captures all the main transport features obtained from the detailed numerical simulations. 
In fact, the upscaled model's predictions are consistent with all the aforementioned properties of the mean displacement and variance time evolutions, as well as those of the BTCs. 
Notably, the CTRW model reproduces well the time evolution of spatial moments from pre-asymptotic to asymptotic transport behavior (with the related scalings, both at early and long times), as well as the BTC tailing observed in the detailed numerical simulations. Its reliance on a few input parameters, i.e., the Eulerian velocity PDF, advective tortuosity, and correlation length, ensures computational efficiency while maintaining accuracy. Together with its reduced computational cost, this makes it a useful tool for upscaled characterization of transport in complex geological fractures, for which fully resolved particle tracking may be impractical. Furthermore, it provides a theoretical explanation of the exponents of the BTCs tailing's power law  from the power law exponents of the Eulerian velocity PDFs at low velocities. Those PDFs arise themselves from the aperture field's variability, which directly links the scaling of late time advective transport to measurable geometric parameters of the fracture.

The transport signatures obtained here are consistent with the broader phenomenology of anomalous transport in heterogeneous porous and fractured media, where velocity heterogeneity, preferential pathways, and low-velocity regions can produce early breakthrough, broad transition-time distributions, and non-Fickian spreading. In geological fractures
Stokes flow 
is channelized and exhibits persistent Lagrangian velocity correlations up to the correlation length. These flow features in turn control plume spreading and breakthrough statistics for advective transport. The ensemble analysis further shows how the representativeness of individual fractures in the fracture population depends on the ratio between correlation length and fracture size, providing a direct assessment of who ergodicity is approached when the former becomes much smaller than the latter. Future work should extend the framework to finite-Péclet-number conditions by including molecular diffusion and assessing how diffusive mixing modifies the impacts of the relative closure and correlation length on solute transport.

\backsection[Supplementary data]{}

\backsection[Acknowledgements]{
A.L. acknowledges that part of the computing for this project was performed on the Sherlock cluster. 
The authors thank Stanford University and the Stanford Research Computing Center for providing computational resources and technical support that contributed to these research results. 
A.L. also acknowledges the Italian SuperComputing Resource Allocation (ISCRA) for granting access to the Leonardo supercomputer, owned by the EuroHPC Joint Undertaking and hosted by CINECA (Italy).}

\backsection[Funding]{
A.L., Y.M., and V.D.F. acknowledge funding from the European Union’s Horizon Europe research and innovation programme under the Marie Skłodowska–Curie grant agreement No.~101111216, Project GEONEAT — “Complex Fluids in Fractured Geological Media for Enhanced Heat Transfer”. 
M.D. acknowledges funding from the European Union’s Horizon Europe research and innovation programme under the European Research Council (ERC) grant agreement No.~101071836, Project KARST — “Karst Aquifer Transport and Reaction Processes from Pore to Regional Scale”. 
A.L. also acknowledges additional support from the Research Fund (RFO) of the Department of Civil, Chemical, Environmental, and Materials Engineering, University of Bologna.

Views and opinions expressed are, however, those of the authors only and do not necessarily reflect those of the European Union or the European Research Council Executive Agency. 
Neither the European Union nor the granting authority can be held responsible for them.}

\backsection[Declaration of interests]{{\bf Declaration of Interests}. The authors report no conflict of interest.}

\backsection[Data availability statement]{
The Monte Carlo datasets generated in this study, including the generated fracture aperture fields, the pressure fields, and the longitudinal and transverse velocity fields obtained by solving the Reynolds equation, are publicly available via Zenodo:

\begin{itemize}
    \item \textbf{Run 01:} \url{https://doi.org/10.5281/zenodo.15252746}
    \item \textbf{Run 02:} \url{https://doi.org/10.5281/zenodo.15253945}
    \item \textbf{Run 03:} \url{https://doi.org/10.5281/zenodo.15256903}
    \item \textbf{Run 04:} \url{https://doi.org/10.5281/zenodo.15257372}
\end{itemize}

This dataset contains 100 Monte Carlo realizations stored in HDF5 format. The datasets are documented in accordance with the GEONEAT Data Management Plan and are compatible with MATLAB, Python (via \texttt{h5py}), and HDFView. Hydrodynamic transport simulations are not included in the shared files due to their large size; however, they can be fully reproduced from the provided aperture and velocity fields using the methods detailed in the manuscript.
}

\backsection[Author ORCIDs]{A. Lenci, https://orcid.org/0000-0002-0285-6991; Y.Méheust, https://orcid.org/0000-0003-1284-3251; M. Dentz, https://orcid.org/0000-0002-3940-282X; V. Di Federico,https://orcid.org/0000-0001-9554-0373}  

\backsection[Author contributions]{
A. Lenci contributed to conceptualization, methodology, formal analysis, software development, validation, investigation, data curation, visualization, writing—original draft, writing—review and editing, and funding acquisition.
Y. Méheust contributed to conceptualization, methodology, formal analysis, writing—review and editing, supervision, and funding acquisition.
M. Dentz contributed to conceptualization, methodology, formal analysis, validation, writing—review and editing, and supervision.
V. Di Federico contributed to conceptualization, supervision, project administration, and funding acquisition. All authors reviewed and approved the final version of the manuscript.
}

\appendix
\section{Time-domain Random Walk Scheme} \label{app:tdrw}
In order to show the equivalence of the TDRW scheme~\eqref{eq:tdrw} and the advection equation~\eqref{eq:advection}, we discretize the spatial derivatives on the right side using an upstream weighting scheme, such that
\begin{align}
\frac{d c^{(i,j)}(t)}{d t} &= - u_1^{(i,j)} \frac{c^{(i,j)}(t) - c^{(i-1,j)}(t)}{\Delta x} H(u_1^{(i,j)}) + u_1^{(i,j)} \frac{c^{(i,j)}(t) - c^{(i+1,j)}(t)}{\Delta x} H(-u_1^{(i,j)})
\nonumber
\\
&- u_2^{(i,j)} \frac{c^{(i,j)}(t) - c^{(i,j-1)}(t)}{\Delta x} H(u_2^{(i,j)}) + u_2^{(i,j)} \frac{c^{(i,j)}(t)- c^{(i,j+1)}(t)}{\Delta x} H(-u_2^{(i,j)}),
\label{app:advection}
\end{align}
where $H(u)$ denotes the Heaviside step function which is one if its argument is positive and zero else. Furthermore, we set
$c^{(i,j)}(t) = c(i \Delta x, j \Delta x,t)$, $u_1^{(i,j)} = u_1(i \Delta x, j\Delta x)$ and $u_2^{(i,j)} = u_2(i \Delta x, j \Delta x)$ at node $(i,j)$ of the regular computational grid. Equation~\eqref{app:advection} can be written as
\begin{align}
\frac{d c^{(i,j)}(t)}{d t}
&= -\,\tau^{(i,j)}\, c^{(i,j)}(t)
+ \tau^{(i,j)} \Big[
   w^{(i,j)}_{\parallel} H\!\big(u_1^{(i,j)}\big)\, c^{(i-1,j)}(t)
 \nonumber\\
&\quad
 + w^{(i,j)}_{\parallel} H\!\big(-u_1^{(i,j)}\big)\, c^{(i+1,j)}(t)
 + w^{(i,j)}_{\perp}  H\!\big(u_2^{(i,j)}\big)\, c^{(i,j-1)}(t)
 \nonumber\\
&\quad
 + w^{(i,j)}_{\perp}  H\!\big(-u_2^{(i,j)}\big)\, c^{(i,j+1)}(t)
 \Big].
\label{app:master}
\end{align}

where we have defined the probabilities to move horizontally or vertically, respectively, as
\begin{align}
w^{(i,j)}_{\parallel} = \frac{|u_1^{(i,j)}|}{|u_1^{(i,j)}| + |u_2^{(i,j)}|} \text{~ and ~} w^{(i,j)}_{\perp} = \frac{|u_2^{(i,j)}|}{|u_1^{(i,j)}| + |u_2^{(i,j)}|} ,
\end{align}
and the transition time at node $(i,j)$ of the regular grid as
\begin{align}
\tau^{(i,j)} = \frac{\Delta x}{|u_1^{(i,j)}| + |u_2^{(i,j)}|}.
\end{align}
The denominator does not represent a physical norm of the velocity vector, but rather serves as a proxy for the total advective flux exiting the cell at node $(i,j)$. Specifically, the sum $|u_1^{(i,j)}| + |u_2^{(i,j)}|$ approximates the total magnitude of the outflow along the principal grid directions, consistent with the upstream weighting discretization. This formulation ensures numerical consistency with the finite-volume fluxes and leads to a stable and efficient random walk approximation of the advection equation. Equation~\eqref{app:master} is a master equation. The first term on the right side describes particle transitions away from node $(i,j)$ during time $\tau^{(i,j)}$, the remaining terms describe downstream horizontal and vertical particle transitions during time $\tau^{(i,j)}$. 

Due to its lattice-based nature, the scheme may be subject to numerical dispersion \citep{Russian2016}, which can be mitigated by adopting a sufficiently fine mesh. The stochastic media generation, along with the flow and transport simulations, was performed using the Sherlock cluster provided by Stanford University and the Stanford Research Computing Center, as well as the Leonardo supercomputer owned by the EuroHPC Joint Undertaking. The datasets were stored in HDF5 format, a hierarchical data format that offers optimal performance in terms of computational time for large arrays. Upscaled model simulations and data post-processing were performed on a standard desktop workstation (Intel(R) Core(TM) i7-4790 CPU @ 3.60 GHz, 16 GB RAM).

\section{Scalings\label{app:scalings}}
The correlated CTRW model presented in Section~\ref{upscale_model} can be coarse-grained on the correlation scale $\ell_\mathrm{c}$. 
For distances larger than $\ell_\mathrm{c}$, successive velocities can be considered statistically independent. 
Particle motion can thus be approximated by the uncorrelated CTRW
\begin{align}
x_1^{(n+1)} &= x_{1}^{(n)} + \frac{\ell_\mathrm{c}}{\chi}, \label{eq:CTRW_step}\\[3pt]
t^{(n+1)} &= t^{(n)} + \tau_n, \qquad 
\tau_n = \frac{\ell_\mathrm{c}}{u_n}, \label{eq:CTRW_time}
\end{align}
where $\chi$ is the advective tortuosity that accounts for streamline curvature. 
Each transition corresponds to a displacement of length $\ell_\mathrm{c}/\chi$ along the mean flow direction and a random travel time $\tau_n$ drawn from the $s$-Lagrangian velocity PDF $f_s(u)$.

The distribution $\psi(t)$ of transition times is given in terms of $f_s(u)$ by
\begin{align}
\psi(t) = \frac{\tau_\textrm{c}}{t^{2}} f_s(\ell_\mathrm{c}/t).  
\end{align}
This implies that
\begin{align}
\psi(t) \propto t^{-2-\alpha} 
\end{align}
for $f_s(u) \propto u^{\alpha}$ or equivalently $f_e(u) \propto u^{\alpha -1}$.
Correspondingly, the transition time distribution $\psi_0(t)$ for the first step is given by
\begin{align}
\psi_0(t) = \frac{\tau_\textrm{c}}{t^{2}} f_0(\ell_\mathrm{c}/t). 
\end{align}
\cite{shlesinger1974} shows that for $0 < \alpha < 1$ the displacement variance scales asymptotically as $\mathcal V(t) \propto t^{2- \alpha}$ and for $\alpha = 1$ as $\mathcal V(t) \propto t \ln t$. 

The late time scaling of the BTCs can also be understood from this reasoning because they scale as the transition time distributions
for an instantaneous injection. For the uniform injection $\psi_0(t) \propto t^{-1-\alpha}$ while $\psi(t) \propto t^{-2-\alpha}$ as indicated above. For uniform injection, the BTC tails are dominated by the long transition times at the first step and thus $\mathcal F(t) \propto t^{-1-\alpha}$. For the flux-weighted injection, $\psi_0(t) = \psi(t)$ and the BTC scales as $\mathcal F(t) \propto t^{-2-\alpha}$. 

\bibliographystyle{jfm}
\bibliography{References}

@article{LBDC2008:2,
	author = {Le Borgne, T. and Dentz, M. and Carrera, J.},
	journal = {Phys. Rev. Lett.},
	pages = {090601},
	title = {A Lagrangian statistical model for transport in highly heterogeneous velocity fields},
	volume = {101},
	year = {2008}}

@article{Berkowitz2006,
	author = {Berkowitz, B. and Cortis, A. and Dentz, M and Scher, H.},
	journal = {Rev. Geophys.},
	pages = {RG2003},
	title = {Modeling non-Fickian transport in geological formations as a continuous time random walk},
	volume = {44},
	year = {2006}}

@article{delay2002,
	author = {Delay, Fr{\'e}d{\'e}rick and Porel, Gilles and Sardini, Paul},
	journal = {Comptes Rendus Geoscience},
	number = {13},
	pages = {967--973},
	publisher = {Elsevier},
	title = {Modelling diffusion in a heterogeneous rock matrix with a time-domain Lagrangian method and an inversion procedure},
	volume = {334},
	year = {2002}}

@article{Cvetkovic2014a,
	author = {Cvetkovic, Vladimir and Fiori, Aldo and Dagan, Gedeon},
	journal = {Water Resources Research},
	number = {7},
	pages = {5759--5773},
	publisher = {Wiley Online Library},
	title = {Solute transport in aquifers of arbitrary variability: A time-domain random walk formulation},
	volume = {50},
	year = {2014}}

@article{dentz2025linear,
        author = {Dentz, M. and Massoudieh, A.},
        journal = {ARC Geophysical Research},
        number = {1},
        publisher = {ARC Alliance},
        title = {Linear Boltzmann Equation for Solute Dispersion in Heterogeneous Media Under Non-Ergodic Conditions},
        volume = {1},
        year = {2025}}

@article{Cvetkovic2014b,
  title={On the upscaling of chemical transport in fractured rock},
  author={Cvetkovic, Vladimir and Gotovac, Hrvoje},
  journal={Water resources research},
  volume={50},
  number={7},
  pages={5797--5816},
  year={2014},
  publisher={Wiley Online Library}
}

@article{Souzy2020,
	author = {M. Souzy and H. Lhuissier and Y. M{\'{e}}heust and T. Le~Borgne and B. Metzger},
	doi = {10.1017/jfm.2020.113},
	journal = {Journal of Fluid Mechanics},
	month = mar,
	publisher = {Cambridge University Press ({CUP})},
	title = {Velocity distributions, dispersion and stretching in three-dimensional porous media},
	url = {https://doi.org/10.1017/jfm.2020.113},
	volume = {891},
	year = {2020}
}

@article{shlesinger1974,
	author = {Shlesinger, Michael F},
	journal = {Journal of Statistical Physics},
	number = {5},
	pages = {421--434},
	publisher = {Springer},
	title = {Asymptotic solutions of continuous-time random walks},
	volume = {10},
	year = {1974}
}

@misc{zenodo_run01,
  author       = {Lenci, Alessandro},
  title        = {GEONEAT 2{D} Heterogeneous Fracture Aperture and {R}eynolds-Based Flow Fields Run 01},
  year         = {2025},
  publisher    = {Zenodo},
  doi          = {10.5281/zenodo.15252746},
  url          = {https://doi.org/10.5281/zenodo.15252746}
}

@misc{zenodo_run02,
  author       = {Lenci, Alessandro},
  title        = {GEONEAT 2{D} Heterogeneous Fracture Aperture and {R}eynolds-Based Flow Fields Run 02},
  year         = {2025},
  publisher    = {Zenodo},
  doi          = {10.5281/zenodo.15253945},
  url          = {https://doi.org/10.5281/zenodo.15253945}
}

@misc{zenodo_run03,
  author       = {Lenci, Alessandro},
  title        = {GEONEAT 2{D} Heterogeneous Fracture Aperture and {R}eynolds-Based Flow Fields Run 03},
  year         = {2025},
  publisher    = {Zenodo},
  doi          = {10.5281/zenodo.15256903},
  url          = {https://doi.org/10.5281/zenodo.15256903}
}

@misc{zenodo_run04,
  author       = {Lenci, Alessandro},
  title        = {GEONEAT 2{D} Heterogeneous Fracture Aperture and Reynolds-Based Flow Fields Run 04},
  year         = {2025},
  publisher    = {Zenodo},
  doi          = {10.5281/zenodo.15257372},
  url          = {https://doi.org/10.5281/zenodo.15257372}
}

@Article{mazzia2011bad,
  author    = {Mazzia, A. and Manzini, G. and Putti, M.},
  journal   = {J. Comput. Phys.},
  title     = {Bad behavior of {G}odunov mixed methods for strongly anisotropic advection--dispersion equations},
  year      = {2011},
  number    = {23},
  pages     = {8410--8426},
  volume    = {230},
  publisher = {Elsevier},
}

@Article{brushWRR2003,
  author    = {Brush, D. J. and Thomson, N. R.},
  journal   = {Water Resour. Res.},
  title     = {Fluid flow in synthetic rough-walled fractures: {N}avier-{S}tokes, {S}tokes, and local cubic law simulations},
  year      = {2003},
  number    = {4},
  volume    = {39},
  publisher = {Wiley Online Library},
}

@Article{Ro-Pl-Hu,
  Title                    = {Tracer dispersion in rough open cracks},
  Author                   = {Roux, S. and Plourabou{\'{e}}, F. and Hulin, J. P.},
  Journal                  = {Transp. Por. Med.},
  Year                     = {1996},
  Number                   = {1},
  Pages                    = {97-116},
  Volume                   = {32}
}

@Article{thompsonJGR1991,
  author    = {Thompson, Mollie E},
  journal   = {J. Geophys. Res. Solid Earth},
  title     = {Numerical simulation of solute transport in rough fractures},
  year      = {1991},
  number    = {B3},
  pages     = {4157--4166},
  volume    = {96},
  publisher = {Wiley Online Library},
}

@Article{drazerPhysRevLett2004,
  Title                    = {Self-affine fronts in self-affine fractures: Large and small-scale structure},
  Author                   = {Drazer, G. and Auradou, H. and Koplik, J. and Hulin, J. P.},
  Journal                  = {Phys. Rev. Lett.},
  Year                     = {2004},
  Number                   = {1},
  Pages                    = {014501},
  Volume                   = {92}
}

@Article{Pl-Hu-Ro-Ko,
  Title                    = {Numerical study of geometrical dispersion in self-affine rough fractures},
  Author                   = {Plourabou{\'{e}}, F. and Hulin, J. P. and Roux, S. and Koplik, J.},
  Journal                  = {Phys. Rev. E},
  Year                     = {1998},
  Number                   = {3},
  Pages                    = {3334-3346},
  Volume                   = {58}
}

@Article{deDreuzyJGR2012,
  author   = {de Dreuzy, J.-R. and M\'eheust, Y. and Pichot, G.},
  journal  = {J. Geophys. Res.: Solid Earth},
  title    = {Influence of fracture scale heterogeneity on the flow properties of three-dimensional discrete fracture networks ({DFN})},
  year     = {2012},
  issn     = {2156-2202},
  number   = {B11207},
  volume   = {117},
  doi      = {10.1029/2012JB009461},
}

@Article{deDreuzyWRR2002,
  author    = {de Dreuzy, J.-R. and Davy, P. and Bour, O.},
  journal   = {Water Resour. Res.},
  title     = {Hydraulic properties of two-dimensional random fracture networks following power law distributions of length and aperture},
  year      = {2002},
  number    = {12},
  pages     = {12--1},
  volume    = {38},
  publisher = {Wiley Online Library},
}

@Article{bourWRR98,
  Title                    = {On the connectivity of three-dimensional fault networks},
  Author                   = {Bour, O. and Davy, P.},
  Journal                  = {Water Resour. Res.},
  Year                     = {1998},
  Number                   = {10},
  Pages                    = {2611-2622},
  Volume                   = {34}
}

@Article{Meheust2003,
  author    = {Y. M{\'{e}}heust and J. Schmittbuhl},
  journal   = {Pure Appl. Geophys.},
  title     = {Scale Effects Related to Flow in Rough Fractures},
  year      = {2003},
  number    = {5},
  pages     = {1023--1050},
  volume    = {160},
  doi       = {10.1007/pl00012559},
  publisher = {Springer Science and Business Media {LLC}},
}

@Article{Neuman2005,
  author    = {S.P. Neuman},
  journal   = {Hydrol. J.},
  title     = {Trends, prospects and challenges in quantifying flow and transport through fractured rocks},
  year      = {2005},
  number    = {1},
  pages     = {124--147},
  volume    = {13},
  doi       = {10.1007/s10040-004-0397-2},
  publisher = {Springer Science and Business Media {LLC}},
}

@Article{Zimmerman1996,
  author    = {R.W. Zimmerman and G.S. Bodvarsson},
  journal   = {Transp. Porous Media},
  title     = {Hydraulic conductivity of rock fractures},
  year      = {1996},
  number    = {1},
  volume    = {23},
  doi       = {10.1007/bf00145263},
  publisher = {Springer Nature},
}

@Article{Brown1987,
  author    = {S. R. Brown},
  journal   = {J. Geophys. Res.},
  title     = {Fluid flow through rock joints: The effect of surface roughness},
  year      = {1987},
  number    = {B2},
  pages     = {1337},
  volume    = {92},
  doi       = {10.1029/jb092ib02p01337},
  publisher = {American Geophysical Union ({AGU})},
}

@Article{Brown1995,
  author    = {S. R. Brown},
  journal   = {J. Geophys. Res. Solid Earth},
  title     = {Simple mathematical model of a rough fracture},
  year      = {1995},
  number    = {B4},
  pages     = {5941--5952},
  volume    = {100},
  doi       = {10.1029/94jb03262},
  publisher = {American Geophysical Union ({AGU})},
}

@Book{Bear1972,
  title     = {Dynamics of Fluids in Porous Media},
  publisher = {Elsevier},
  year      = {1972},
  author    = {J. Bear},
}

@Article{Meheust2001,
  author    = {Y. M{\'{e}}heust and J. Schmittbuhl},
  journal   = {J. Geophys. Res. Solid Earth},
  title     = {Geometrical heterogeneities and permeability anisotropy of rough fractures},
  year      = {2001},
  number    = {B2},
  pages     = {2089--2102},
  volume    = {106},
  doi       = {10.1029/2000jb900306},
  publisher = {American Geophysical Union ({AGU})},
}

@Article{Cardenas2007,
  author  = {M. B. Cardenas and D. T. Slottke and R. A. Ketcham and J. M. Sharp},
  journal = {Geophys. Res. Lett.},
  title   = {{N}avier-{S}tokes flow and transport simulations using real fractures shows heavy tailing due to eddies},
  year    = {2007},
  number  = {L14404},
  volume  = {34},
  doi     = {10.1029/2007GL030545},
}

@Article{Comolli2019,
  author    = {A. Comolli and V. Hakoun and M. Dentz},
  journal   = {Water Resour. Res.},
  title     = {Mechanisms, Upscaling, and Prediction of Anomalous Dispersion in Heterogeneous Porous Media},
  year      = {2019},
  number    = {10},
  pages     = {8197--8222},
  volume    = {55},
  doi       = {10.1029/2019wr024919},
  publisher = {American Geophysical Union ({AGU})},
}

@Article{Bouchaud1997,
  author    = {E. Bouchaud},
  journal   = {J. Phys.: Condens. Matter},
  title     = {Scaling properties of cracks},
  year      = {1997},
  number    = {21},
  pages     = {4319--4344},
  volume    = {9},
  doi       = {10.1088/0953-8984/9/21/002},
  publisher = {{IOP} Publishing},
}

@Article{Candela2009,
  author    = {T. Candela and F. Renard and M. Bouchon and A. Brouste and D. Marsan and J. Schmittbuhl and C. Voisin},
  journal   = {Pure Appl. Geophys.},
  title     = {Characterization of Fault Roughness at Various Scales: Implications of Three-Dimensional High Resolution Topography Measurements},
  year      = {2009},
  number    = {10-11},
  pages     = {1817--1851},
  volume    = {166},
  doi       = {10.1007/s00024-009-0521-2},
  publisher = {Springer Science and Business Media {LLC}},
}

@Article{Boffa1999,
  author    = {J. M. Boffa and C. Allain and R. Chertcoff and J. P. Hulin and F. Plourabou{\'{e}} and S. Roux},
  journal   = {Eur. Phys. J. B},
  title     = {Roughness of sandstone fracture surfaces: Profilometry and shadow length investigations},
  year      = {1999},
  number    = {2},
  pages     = {179--182},
  volume    = {7},
  doi       = {10.1007/s100510050602},
  publisher = {Springer Science and Business Media {LLC}},
}

@Article{Dentz2016,
  author    = {M. Dentz and P. K. Kang and A. Comolli and T. Le Borgne and D. R. Lester},
  journal   = {Phys. Rev. Fluids},
  title     = {Continuous time random walks for the evolution of {L}agrangian velocities},
  year      = {2016},
  number    = {7},
  pages     = {074004},
  volume    = {1},
  doi       = {10.1103/physrevfluids.1.074004},
  publisher = {American Physical Society ({APS})},
}

@Article{Puyguiraud2019b,
  author    = {A. Puyguiraud and P. Gouze and M. Dentz},
  journal   = {Transp. Porous Media},
  title     = {Upscaling of Anomalous Pore-Scale Dispersion},
  year      = {2019},
  number    = {2},
  pages     = {837--855},
  volume    = {128},
  doi       = {10.1007/s11242-019-01273-3},
  publisher = {Springer Science and Business Media {LLC}},
}

@Article{Hakoun2019,
  author    = {V. Hakoun and A. Comolli and M. Dentz},
  journal   = {Water Resour. Res.},
  title     = {Upscaling and Prediction of {L}agrangian Velocity Dynamics in Heterogeneous Porous Media},
  year      = {2019},
  number    = {5},
  pages     = {3976--3996},
  volume    = {55},
  doi       = {10.1029/2018wr023810},
  publisher = {American Geophysical Union ({AGU})},
}

@Article{Bouchaud1990,
  author    = {E. Bouchaud and G. Lapasset and J. Plan{\`{e}}s},
  journal   = {Europhysics Letters ({EPL})},
  title     = {Fractal Dimension of Fractured Surfaces: A Universal Value?},
  year      = {1990},
  number    = {1},
  pages     = {73--79},
  volume    = {13},
  doi       = {10.1209/0295-5075/13/1/013},
  publisher = {{IOP} Publishing},
}

@Article{Schmittbuhl1995a,
  author    = {J. Schmittbuhl and F. Schmitt and C. Scholz},
  journal   = {J. Geophys. Res. Solid Earth},
  title     = {Scaling invariance of crack surfaces},
  year      = {1995},
  number    = {B4},
  pages     = {5953--5973},
  volume    = {100},
  doi       = {10.1029/94jb02885},
  publisher = {American Geophysical Union ({AGU})},
}

@Article{Gale2014,
  author    = {J.F.W. Gale and S.E. Laubach and J.E. Olson and P. Eichhuble and A. Fall},
  journal   = {{AAPG} Bulletin},
  title     = {Natural Fractures in shale: A review and new observations},
  year      = {2014},
  number    = {11},
  pages     = {2165--2216},
  volume    = {98},
  doi       = {10.1306/08121413151},
  publisher = {American Association of Petroleum Geologists {AAPG}/Datapages},
}

@Article{Lenci2022b,
  author    = {A. Lenci and Y. M{\'{e}}heust and M. Putti and V. {Di Federico}},
  journal   = {Water Resources Res.},
  title     = {Monte {C}arlo Simulations of Shear-Thinning Flow in Geological Fractures},
  year      = {2022},
  number    = {9},
  volume    = {58},
  doi       = {10.1029/2022wr032024},
  fjournal  = {Water Resources Research},
  publisher = {American Geophysical Union ({AGU})},
}

@Article{Lenci2022a,
  author    = {A. Lenci and M. Putti and V. {Di Federico} and Y. M{\'{e}}heust},
  journal   = {Water Resources Res.},
  title     = {A Lubrication-Based Solver for Shear-Thinning Flow in Rough Fractures},
  year      = {2022},
  number    = {8},
  volume    = {58},
  doi       = {10.1029/2021wr031760},
  fjournal  = {Water Resources Research},
  publisher = {American Geophysical Union ({AGU})},
}

@Article{Meheust2000,
  author    = {Y. M{\'{e}}heust and J. Schmittbuhl},
  journal   = {Geophys. Res. Lett.},
  title     = {Flow enhancement of a rough fracture},
  year      = {2000},
  number    = {18},
  pages     = {2989--2992},
  volume    = {27},
  doi       = {10.1029/1999gl008464},
  publisher = {American Geophysical Union ({AGU})},
}

@Article{Pollock1988,
  author    = {Pollock, D.W.},
  journal   = {Groundwater},
  title     = {Semianalytical Computation of Path Lines for Finite‐Difference Models},
  year      = {1988},
  issn      = {1745-6584},
  month     = nov,
  number    = {6},
  pages     = {743--750},
  volume    = {26},
  doi       = {10.1111/j.1745-6584.1988.tb00425.x},
  publisher = {Wiley},
}

@Article{Kang2017,
  author    = {Kang, P.K. and Dentz, M. and Le Borgne, T. and Lee, S. and Juanes, R.},
  journal   = {Adv. Water Resour.},
  title     = {Anomalous transport in disordered fracture networks: Spatial {M}arkov model for dispersion with variable injection modes},
  year      = {2017},
  issn      = {0309-1708},
  pages     = {80--94},
  volume    = {106},
  doi       = {10.1016/j.advwatres.2017.03.024},
  fjournal  = {Advances in Water Resources},
  publisher = {Elsevier BV},
}

@Article{Morales2017,
  author    = {Morales, V. L. and Dentz, M. and Willmann, M. and Holzner, M.},
  journal   = {Geophys. Res. Lett.},
  title     = {Stochastic dynamics of intermittent pore‐scale particle motion in three‐dimensional porous media: Experiments and theory},
  year      = {2017},
  issn      = {1944-8007},
  number    = {18},
  pages     = {9361--9371},
  volume    = {44},
  doi       = {10.1002/2017gl074326},
  fjournal  = {Geophysical Research Letters},
  publisher = {American Geophysical Union (AGU)},
}

@Article{Gelhar1992,
  author    = {Gelhar, L.W. and Welty, C. and Rehfeldt, K.R.},
  journal   = {Water Resour. Res.},
  title     = {A critical review of data on field‐scale dispersion in aquifers},
  year      = {1992},
  issn      = {1944-7973},
  month     = jul,
  number    = {7},
  pages     = {1955--1974},
  volume    = {28},
  doi       = {10.1029/92wr00607},
  fjournal  = {Water Resources Research},
  publisher = {American Geophysical Union (AGU)},
}

@Article{Silliman1987,
  author    = {Silliman, S.E. and Simpson, E.S.},
  journal   = {Water Resour. Res.},
  title     = {Laboratory evidence of the scale effect in dispersion of solutes in porous media},
  year      = {1987},
  issn      = {1944-7973},
  month     = aug,
  number    = {8},
  pages     = {1667--1673},
  volume    = {23},
  doi       = {10.1029/wr023i008p01667},
  fjournal  = {Water Resources Research},
  publisher = {American Geophysical Union (AGU)},
}

@Article{Warren1964,
  author    = {Warren, J.E. and Skiba, F.F.},
  journal   = {Soc. Petrol. Eng. J.},
  title     = {Macroscopic Dispersion},
  year      = {1964},
  issn      = {0197-7520},
  month     = sep,
  number    = {03},
  pages     = {215--230},
  volume    = {4},
  doi       = {10.2118/648-pa},
  fjournal  = {Society of Petroleum Engineers Journal},
  publisher = {Society of Petroleum Engineers (SPE)},
}

@Article{Pickens1981,
  author    = {Pickens, J.F. and Grisak, G.E.},
  journal   = {Water Resour. Res.},
  title     = {Modeling of scale‐dependent dispersion in hydrogeologic systems},
  year      = {1981},
  issn      = {1944-7973},
  month     = dec,
  number    = {6},
  pages     = {1701--1711},
  volume    = {17},
  doi       = {10.1029/wr017i006p01701},
  fjournal  = {Water Resources Research},
  publisher = {American Geophysical Union (AGU)},
}

@Article{Wang2014,
  author    = {Wang, L. and Cardenas, M. B.},
  journal   = {Water Resour. Res.},
  title     = {Non-Fickian transport through two-dimensional rough fractures: Assessment and prediction},
  year      = {2014},
  issn      = {0043-1397},
  month     = feb,
  number    = {2},
  pages     = {871--884},
  volume    = {50},
  doi       = {10.1002/2013wr014459},
  fjournal  = {Water Resources Research},
  publisher = {American Geophysical Union (AGU)},
}

@Article{Hyman2015,
  author    = {Hyman, J.D. and Painter, S. L. and Viswanathan, H. and Makedonska, N. and Karra, S.},
  journal   = {Water Resour. Res.},
  title     = {Influence of injection mode on transport properties in kilometer-scale three-dimensional discrete fracture networks},
  year      = {2015},
  issn      = {0043-1397},
  month     = sep,
  number    = {9},
  pages     = {7289--7308},
  volume    = {51},
  doi       = {10.1002/2015wr017151},
  fjournal  = {Water Resources Research},
  publisher = {American Geophysical Union (AGU)},
}

@Article{Saffman1959,
  author    = {Saffman, P. G.},
  journal   = {J. Fluid Mech.},
  title     = {A theory of dispersion in a porous medium},
  year      = {1959},
  issn      = {1469-7645},
  month     = oct,
  number    = {03},
  pages     = {321},
  volume    = {6},
  doi       = {10.1017/s0022112059000672},
  fjournal  = {Journal of Fluid Mechanics},
  publisher = {Cambridge University Press (CUP)},
}

@Article{deJosselin1958,
  author    = {De Josselin De Jong, G.},
  journal   = {Eos, Transactions American Geophysical Union},
  title     = {Longitudinal and transverse diffusion in granular deposits},
  year      = {1958},
  issn      = {0002-8606},
  number    = {1},
  pages     = {67--74},
  volume    = {39},
  doi       = {10.1029/tr039i001p00067},
  publisher = {American Geophysical Union (AGU)},
}

@Article{Tran2021,
  author    = {Tran, E. and Zavrin, M. and Kersting, A.B. and Klein-BenDavid, O. and Teutsch, N. and Weisbrod, N.},
  journal   = {Sci. Total Environ.},
  title     = {Colloid-facilitated transport of 238{P}u, 233{u} and 137{C}s through fractured chalk: Laboratory experiments, modelling, and implications for nuclear waste disposal},
  year      = {2021},
  issn      = {0048-9697},
  month     = feb,
  pages     = {143818},
  volume    = {757},
  doi       = {10.1016/j.scitotenv.2020.143818},
  fjournal  = {Science of The Total Environment},
  publisher = {Elsevier BV},
}

@Article{Yoo2021,
  author    = {Yoo, H. and Park, S. and Xie, L. and Kim, K.-I. and Min, K.-B. and Rutqvist, J. and Rinaldi, A.P.},
  journal   = {Geothermics},
  title     = {Hydro-mechanical modeling of the first and second hydraulic stimulations in a fractured geothermal reservoir in Pohang, South Korea},
  year      = {2021},
  issn      = {0375-6505},
  month     = jan,
  pages     = {101982},
  volume    = {89},
  doi       = {10.1016/j.geothermics.2020.101982},
  publisher = {Elsevier BV},
}

@Article{Wang2023,
  author    = {Wang, Z. and Li, H. and Liu, S. and Xu, J. and Liu, J. and Wang, X.},
  journal   = {Fuel},
  title     = {Risk evaluation of {CO} 2 leakage through fracture zone in geological storage reservoir},
  year      = {2023},
  issn      = {0016-2361},
  month     = jun,
  pages     = {127896},
  volume    = {342},
  doi       = {10.1016/j.fuel.2023.127896},
  publisher = {Elsevier BV},
}

@Article{SanchezVila2006,
  author    = {Sanchez‐Vila, X. and Guadagnini, A. and Carrera, J.},
  journal   = {Rev. Geophys.},
  title     = {Representative hydraulic conductivities in saturated groundwater flow},
  year      = {2006},
  issn      = {1944-9208},
  month     = sep,
  number    = {3},
  volume    = {44},
  doi       = {10.1029/2005rg000169},
  fjournal  = {Reviews of Geophysics},
  publisher = {American Geophysical Union (AGU)},
}

@Article{Dentz2023,
  author    = {Dentz, M. and Hyman, J. D.},
  journal   = {J. Fluid Mech.},
  title     = {The hidden structure of hydrodynamic transport in random fracture networks},
  year      = {2023},
  issn      = {1469-7645},
  month     = dec,
  volume    = {977},
  doi       = {10.1017/jfm.2023.973},
  fjournal  = {Journal of Fluid Mechanics},
  publisher = {Cambridge University Press (CUP)},
}

@Article{Hadgu2017,
  author    = {Hadgu, T. and Karra, S. and Kalinina, E. and Makedonska, N. and Hyman, J.D. and Klise, K. and Viswanathan, H.S. and Wang, Y.},
  journal   = {J. Hydrol.},
  title     = {A comparative study of discrete fracture network and equivalent continuum models for simulating flow and transport in the far field of a hypothetical nuclear waste repository in crystalline host rock},
  year      = {2017},
  issn      = {0022-1694},
  month     = oct,
  pages     = {59--70},
  volume    = {553},
  doi       = {10.1016/j.jhydrol.2017.07.046},
  fjournal  = {Journal of Hydrology},
  publisher = {Elsevier BV},
}

@Article{Noetinger2016,
  author    = {Noetinger, B. and Roubinet, D. and Russian, A. and Le Borgne, T. and Delay, F. and Dentz, M. and de Dreuzy, J.-R. and Gouze, P.},
  journal   = {Transport Porous Med.},
  title     = {Random Walk Methods for Modeling Hydrodynamic Transport in Porous and Fractured Media from Pore to Reservoir Scale},
  year      = {2016},
  issn      = {1573-1634},
  month     = apr,
  number    = {2},
  pages     = {345--385},
  volume    = {115},
  doi       = {10.1007/s11242-016-0693-z},
  fjournal  = {Transport in Porous Media},
  publisher = {Springer Science and Business Media LLC},
}

@Article{Berkowitz1997,
  author    = {Berkowitz, B. and Scher, H.},
  journal   = {Phys. Rev. Lett.},
  title     = {Anomalous Transport in Random Fracture Networks},
  year      = {1997},
  issn      = {1079-7114},
  month     = nov,
  number    = {20},
  pages     = {4038--4041},
  volume    = {79},
  doi       = {10.1103/physrevlett.79.4038},
  fjournal  = {Physical Review Letters},
  publisher = {American Physical Society (APS)},
}

@Article{Benke2003,
  author    = {Benke, R. and Painter, S.},
  journal   = {Water Resour. Res.},
  title     = {Modeling conservative tracer transport in fracture networks with a hybrid approach based on the {B}oltzmann transport equation},
  year      = {2003},
  issn      = {1944-7973},
  month     = nov,
  number    = {11},
  volume    = {39},
  doi       = {10.1029/2003wr001966},
  fjournal  = {Water Resources Research},
  publisher = {American Geophysical Union (AGU)},
}

@Article{Kang2011,
  author    = {Kang, P.K. and Dentz, M. and Le Borgne, T. and Juanes, R.},
  journal   = {Phys. Rev. Lett.},
  title     = {Spatial {M}arkov Model of Anomalous Transport Through Random Lattice Networks},
  year      = {2011},
  issn      = {1079-7114},
  month     = oct,
  number    = {18},
  pages     = {180602},
  volume    = {107},
  doi       = {10.1103/physrevlett.107.180602},
  fjournal  = {Physical Review Letters},
  publisher = {American Physical Society (APS)},
}

@Book{Dagan1989,
  author    = {Dagan, G.},
  publisher = {Springer},
  title     = {Flow and transport in porous formations},
  year      = {1989},
  address   = {Berlin},
  isbn      = {0387516026},
  note      = {Literaturverz. S. 451 - 461},
  pagetotal = {465},
  ppn_gvk   = {275084264},
}

@Book{Gelhar1993,
  author    = {Gelhar, L.W.},
  publisher = {Prentice-Hall},
  title     = {Stochastic subsurface hydrology},
  year      = {1993},
  address   = {Englewood Cliffs, NJ},
  isbn      = {0138467676},
  pagetotal = {390},
  ppn_gvk   = {27774797X},
}

@Book{Poinssot2012,
  author    = {Poinssot, C.},
  publisher = {Elsevier Science \& Technology},
  title     = {Radionuclide Behaviour in the Natural Environment},
  year      = {2012},
  address   = {Cambridge},
  isbn      = {9780857097194},
  number    = {42},
  series    = {Woodhead publishing series in energy},
  pagetotal = {1737},
  ppn_gvk   = {1818879336},
  subtitle  = {Science, Implications and Lessons for the Nuclear Industry},
}

@Book{Rubin2003,
  author    = {Rubin, Y.},
  publisher = {Oxford University Press},
  title     = {Applied Stochastic Hydrogeology},
  year      = {2003},
  isbn      = {9780197561676},
  month     = apr,
  doi       = {10.1093/oso/9780195138047.001.0001},
}

@Article{Dagan1984,
  author    = {Dagan, G.},
  journal   = {J. Fluid Mech.},
  title     = {Solute transport in heterogeneous porous formations},
  year      = {1984},
  issn      = {1469-7645},
  month     = aug,
  number    = {1},
  pages     = {151},
  volume    = {145},
  doi       = {10.1017/s0022112084002858},
  fjournal  = {Journal of Fluid Mechanics},
  publisher = {Cambridge University Press (CUP)},
}

@Article{Gotovac2009,
  author    = {Gotovac, Hrvoje and Cvetkovic, Vladimir and Andricevic, Roko},
  journal   = {Water Resour. Res.},
  title     = {Flow and travel time statistics in highly heterogeneous porous media},
  year      = {2009},
  issn      = {1944-7973},
  month     = jul,
  number    = {7},
  volume    = {45},
  doi       = {10.1029/2008wr007168},
  fjournal  = {Water Resources Research},
  publisher = {American Geophysical Union (AGU)},
}

@Article{Long1985,
  author    = {Long, J.C.S. and Witherspoon, P.A.},
  journal   = {J. Geophys. Res. Solid Earth},
  title     = {The relationship of the degree of interconnection to permeability in fracture networks},
  year      = {1985},
  issn      = {0148-0227},
  month     = mar,
  number    = {B4},
  pages     = {3087--3098},
  volume    = {90},
  doi       = {10.1029/jb090ib04p03087},
  publisher = {American Geophysical Union (AGU)},
}

@book{Yeh_Khaleel_Carroll_2015,
place={Cambridge},
title={Flow through Heterogeneous Geologic Media},
publisher={Cambridge University Press},
author={Yeh, T.-C. and Khaleel, R. and Carroll, K.C.},
year={2015},
}

@Article{DiFederico1997,
  author   = {Di Federico, V. and Neuman, S.P.},
  journal  = {Water Resour. Res.},
  title    = {Scaling of random fields by means of truncated power variograms and associated spectra},
  year     = {1997},
  number   = {5},
  pages    = {1075-1085},
  volume   = {33},
  doi      = {https://doi.org/10.1029/97WR00299},
  fjournal = {Water Resources Research},
}

@Article{DiFederico1998a,
  author   = {Di Federico, V. and Neuman, S.P.},
  journal  = {Water Resour. Res.},
  title    = {Flow in multiscale log conductivity fields with truncated power variograms},
  year     = {1998},
  number   = {5},
  pages    = {975-987},
  volume   = {34},
  doi      = {https://doi.org/10.1029/98WR00220},
  fjournal = {Water Resources Research},
}

@Article{DiFederico1998b,
  author   = {Di Federico, V. and Neuman, S.P.},
  journal  = {Water Resour. Res.},
  title    = {Transport in multiscale log conductivity fields with truncated power variograms},
  year     = {1998},
  number   = {5},
  pages    = {963-973},
  volume   = {34},
  doi      = {https://doi.org/10.1029/98WR00221},
  fjournal = {Water Resources Research},
}

@InBook{Jing2007,
  author    = {Jing, L. and Stephansson, O.},
  pages     = {111--144},
  publisher = {Elsevier},
  title     = {Fluid Flow and Coupled Hydro-Mechanical Behavior of Rock Fractures},
  year      = {2007},
  booktitle = {Developments in Geotechnical Engineering},
  doi       = {10.1016/s0165-1250(07)85004-8},
  issn      = {0165-1250},
}

@Article{Watanabe2008,
  author   = {Watanabe, N. and Hirano, N. and Tsuchiya, N.},
  journal  = {Water Resour. Res.},
  title    = {Determination of aperture structure and fluid flow in a rock fracture by high-resolution numerical modeling on the basis of a flow-through experiment under confining pressure},
  year     = {2008},
  number   = {6},
  volume   = {44},
  doi      = {https://doi.org/10.1029/2006WR005411},
  fjournal = {Water Resources Research},
}

@Article{Fiori2017,
  author               = {Fiori, A. and Zarlenga, A. and Jankovic, I. and Dagan, G.},
  journal              = {Adv. Water Resour.},
  title                = {Solute transport in aquifers: The comeback of the advection dispersion equation and the First Order Approximation},
  year                 = {2017},
  issn                 = {0309-1708},
  month                = {DEC},
  pages                = {349-359},
  volume               = {110},
  doi                  = {10.1016/j.advwatres.2017.10.025},
  eissn                = {1872-9657},
  fjournal             = {Advances in Water Resources},
  orcid-numbers        = {Fiori, Aldo/0000-0002-6662-5738 Zarlenga, Antonio/0000-0001-7043-4718},
  researcherid-numbers = {JANKOVIC, IGOR/E-5585-2015 Fiori, Aldo/A-2321-2010 Zarlenga, Antonio/D-3971-2018},
  unique-id            = {WOS:000418262400027},
}

@Article{Zech2018,
  author               = {Zech, A. and D'Angelo, C. and Attinger, S. and Fiori, A.},
  journal              = {Adv. Water Resour.},
  title                = {Revisitation of the dipole tracer test for heterogeneous porous formations},
  year                 = {2018},
  issn                 = {0309-1708},
  month                = {MAY},
  pages                = {198-206},
  volume               = {115},
  doi                  = {10.1016/j.advwatres.2018.03.006},
  eissn                = {1872-9657},
  fjournal             = {Advances in Water Resources},
  orcid-numbers        = {Zech, Alraune/0000-0002-8783-6198 Fiori, Aldo/0000-0002-6662-5738},
  researcherid-numbers = {Zech, Alraune/P-6232-2019 Fiori, Aldo/A-2321-2010},
  unique-id            = {WOS:000432554300015},
}

@Article{Russian2016,
  author               = {Russian, A. and Dentz, M. and Gouze, P.},
  journal              = {Water Resour. Res.},
  title                = {Time domain random walks for hydrodynamic transport in heterogeneous media},
  year                 = {2016},
  issn                 = {0043-1397},
  month                = {MAY},
  number               = {5},
  pages                = {3309-3323},
  volume               = {52},
  doi                  = {10.1002/2015WR018511},
  eissn                = {1944-7973},
  fjournal             = {Water Resources Research},
  orcid-numbers        = {Dentz, Marco/0000-0002-3940-282X Gouze, Philippe/0000-0002-8912-5453},
  researcherid-numbers = {Gouze, Philippe/GQA-9847-2022 Dentz, Marco/C-1076-2015 Gouze, Philippe/A-3929-2010 Gouze, Philippe/D-5040-2011},
  unique-id            = {WOS:000379259800003},
}

@Article{Lenci2024,
  author    = {Lenci, A. and Méheust, Y. and Di Federico, V. and Ciriello, V.},
  journal   = {Water Resour. Res.},
  title     = {Reduced‐Order Models Unravel the Joint Impact of Aperture Heterogeneity and Shear‐Thinning Rheology on Fracture‐Scale Flow Metrics},
  year      = {2024},
  issn      = {1944-7973},
  month     = jan,
  number    = {1},
  volume    = {60},
  doi       = {10.1029/2023wr036026},
  fjournal  = {Water Resources Research},
  publisher = {American Geophysical Union (AGU)},
}

@Book{Gardiner2009,
  author    = {Gardiner, C. W.},
  publisher = {Springer},
  title     = {Stochastic methods},
  year      = {2009},
  address   = {Berlin},
  edition   = {4th ed.},
  isbn      = {9783642089626},
  note      = {Includes bibliographical references and index. - Previous ed.: 2004},
  number    = {1},
  series    = {Springer complexity},
  pagetotal = {447},
  ppn_gvk   = {1617632848},
  subtitle  = {A handbook for the natural and social sciences},
}

@Article{Viswanathan2022,
  author    = {Viswanathan, H.S. and Ajo‐Franklin, J. and Birkholzer, J.T. and Carey, J.W. and Guglielmi, Y. and Hyman, J.D. and Karra, S. and Pyrak‐Nolte, L.J. and Rajaram, H. and Srinivasan, G. and Tartakovsky, D.M.},
  journal   = {Rev. Geophys.},
  title     = {From Fluid Flow to Coupled Processes in Fractured Rock: Recent Advances and New Frontiers},
  year      = {2022},
  issn      = {1944-9208},
  month     = feb,
  number    = {1},
  volume    = {60},
  doi       = {10.1029/2021rg000744},
  publisher = {American Geophysical Union (AGU)},
}

@Article{deAnna2017,
  author   = {de Anna, P. and Quaife, B. and Biros, G. and Juanes, R.},
  journal  = {Phys. Rev. Fluids},
  title    = {Prediction of the low-velocity distribution from the pore structure in simple porous media},
  year     = {2017},
  number   = {12},
  pages    = {124103},
  volume   = {2},
  doi      = {10.1103/physrevfluids.2.124103},
  fjournal = {Physical Review Fluids},
}

@Article{Hyman2021,
  author   = {Hyman, J.D. and Dentz, M.},
  journal  = {Adv. Water Resour.},
  title    = {Transport Upscaling under Flow Heterogeneity and Matrix-Diffusion in Three-Dimensional Discrete Fracture Networks},
  year     = {2021},
  pages    = {103994},
  volume   = {155},
  doi      = {10.1016/j.advwatres.2021.103994},
  fjournal = {Advances in Water Resources},
}

@Article{Dentz2004,
  author   = {Dentz, M. and Cortis, A. and Scher, H. and Berkowitz, B.},
  journal  = {Adv. Water Resour.},
  title    = {Time behavior of solute transport in heterogeneous media: transition from anomalous to normal transport},
  year     = {2004},
  number   = {2},
  pages    = {155--173},
  volume   = {27},
  doi      = {10.1016/j.advwatres.2003.11.002},
  fjournal = {Advances in Water Resources},
}

@Article{Andrade1997,
  author   = {Andrade, J. S. and Almeida, M. P. and Mendes Filho, J. and Havlin, S. and Suki, B. and Stanley, H. E.},
  journal  = {Phys. Rev. Lett.},
  title    = {Fluid Flow through Porous Media: The Role of Stagnant Zones},
  year     = {1997},
  number   = {20},
  pages    = {3901--3904},
  volume   = {79},
  doi      = {10.1103/physrevlett.79.3901},
  fjournal = {Physical Review Letters},
}

@Article{Tyukhova2016,
  author   = {Tyukhova, A. and Dentz, M. and Kinzelbach, W. and Willmann, M.},
  journal  = {Phys. Rev. Fluids},
  title    = {Mechanisms of anomalous dispersion in flow through heterogeneous porous media},
  year     = {2016},
  number   = {7},
  pages    = {074002},
  volume   = {1},
  doi      = {10.1103/physrevfluids.1.074002},
  fjournal = {Physical Review Fluids},
}

@Article{Zvikelsky2006,
  author    = {Zvikelsky, O. and Weisbrod, N.},
  journal   = {Water Resources Research},
  title     = {Impact of particle size on colloid transport in discrete fractures},
  year      = {2006},
  issn      = {1944-7973},
  month     = nov,
  number    = {12},
  volume    = {42},
  doi       = {10.1029/2006wr004873},
  publisher = {American Geophysical Union (AGU)},
}

@Article{Dontsov2014,
  author    = {Dontsov, E. V. and Peirce, A. P.},
  journal   = {Journal of Fluid Mechanics},
  title     = {Slurry flow, gravitational settling and a proppant transport model for hydraulic fractures},
  year      = {2014},
  issn      = {1469-7645},
  month     = nov,
  pages     = {567--590},
  volume    = {760},
  doi       = {10.1017/jfm.2014.606},
  publisher = {Cambridge University Press (CUP)},
}

@Article{Cheng2025,
  author    = {Cheng, S. and Wu, B. and Huppert, H. E. and Ma, T. and Chen, Z. and Tan, P.},
  journal   = {Journal of Rock Mechanics and Geotechnical Engineering},
  title     = {Modeling transport and bridging behavior of lost circulation materials in a hydraulic fracture},
  year      = {2025},
  issn      = {1674-7755},
  month     = mar,
  doi       = {10.1016/j.jrmge.2025.03.005},
  publisher = {Elsevier BV},
}

@Article{Hyman2019,
  author    = {Hyman, J. D. and Dentz, M. and Hagberg, A. and Kang, P. K.},
  journal   = {Journal of Geophysical Research: Solid Earth},
  title     = {Linking Structural and Transport Properties in Three‐Dimensional Fracture Networks},
  year      = {2019},
  issn      = {2169-9356},
  month     = feb,
  number    = {2},
  pages     = {1185--1204},
  volume    = {124},
  doi       = {10.1029/2018jb016553},
  publisher = {American Geophysical Union (AGU)},
}

@Article{Edery2016,
  author    = {Edery, Y. and Geiger, S. and Berkowitz, B.},
  journal   = {Water Resources Research},
  title     = {Structural controls on anomalous transport in fractured porous rock},
  year      = {2016},
  issn      = {1944-7973},
  month     = jul,
  number    = {7},
  pages     = {5634--5643},
  volume    = {52},
  doi       = {10.1002/2016wr018942},
  publisher = {American Geophysical Union (AGU)},
}

@Article{Sund2021,
  author    = {Sund, Nicole L. and Parashar, Rishi and Pham, Hai V.},
  journal   = {Physical Review E},
  title     = {Upscaling of transport through discrete fracture networks via random walk: A comparison of models},
  year      = {2021},
  issn      = {2470-0053},
  month     = jun,
  number    = {6},
  pages     = {062116},
  volume    = {103},
  doi       = {10.1103/physreve.103.062116},
  publisher = {American Physical Society (APS)},
}

@Article{Elhanati2024,
  author    = {Elhanati, D. and Goeppert, N. and Berkowitz, B.},
  journal   = {Hydrology and Earth System Sciences},
  title     = {Karst aquifer discharge response to rainfall interpreted as anomalous transport},
  year      = {2024},
  issn      = {1607-7938},
  month     = sep,
  number    = {17},
  pages     = {4239--4249},
  volume    = {28},
  doi       = {10.5194/hess-28-4239-2024},
  publisher = {Copernicus GmbH},
}

@Article{Fiori2015,
  author    = {Fiori, A. and Becker, M.W.},
  journal   = {Journal of Hydrology},
  title     = {Power law breakthrough curve tailing in a fracture: The role of advection},
  year      = {2015},
  issn      = {0022-1694},
  month     = jun,
  pages     = {706--710},
  volume    = {525},
  doi       = {10.1016/j.jhydrol.2015.04.029},
  publisher = {Elsevier BV},
}

@Article{Painter2002,
  author    = {Painter, Scott and Cvetkovic, Vladimir and Selroos, Jan‐Olof},
  journal   = {Geophysical Research Letters},
  title     = {Power‐law velocity distributions in fracture networks: Numerical evidence and implications for tracer transport},
  year      = {2002},
  issn      = {1944-8007},
  month     = jul,
  number    = {14},
  volume    = {29},
  doi       = {10.1029/2002gl014960},
  publisher = {American Geophysical Union (AGU)},
}

@Article{Fiori2007,
  author    = {Fiori, Aldo and Janković, Igor and Dagan, Gedeon and Cvetković, Vladimir},
  journal   = {Water Resources Research},
  title     = {Ergodic transport through aquifers of non‐Gaussian log conductivity distribution and occurrence of anomalous behavior},
  year      = {2007},
  issn      = {1944-7973},
  month     = sep,
  number    = {9},
  volume    = {43},
  doi       = {10.1029/2007wr005976},
  publisher = {American Geophysical Union (AGU)},
}
\end{document}